\documentclass{aastex631}

\newcommand\tde{AT2018hyz}

\usepackage{amsmath,amstext}
\usepackage[T1]{fontenc}
\usepackage[figure,figure*]{hypcap}
\usepackage{graphicx}
\usepackage{subfigure}
\usepackage{tabularx}
\usepackage{enumitem}
\usepackage{appendix}

\shorttitle{Late-time Radio Emission in AT2018hyz}
\shortauthors{Cendes et al.}

\graphicspath{{./}{figures/}}

\begin{document}


\title[AT2018hyz]{Continued Rapid Radio Brightening of the Tidal Disruption Event AT2018hyz}


\correspondingauthor{Yvette Cendes}
\email{yncendes@uoregon.edu}

\author[0000-0001-7007-6295]{Yvette Cendes}
\affiliation{Department of Physics, University of Oregon, 1371 E 13th Ave, Eugene OR 97403, USA}
\affiliation{Institute for Fundamental Science, University of Oregon, 1371 E 13th Ave, Eugene OR 97403, USA}

\author[0000-0002-9392-9681]{Edo Berger}
\affiliation{Center for Astrophysics | Harvard \& Smithsonian, Cambridge, MA 02138, USA}

\author[0000-0001-7833-1043]{Paz Beniamini}
\affiliation{Department of Natural Sciences, The Open University of Israel, P.O Box 808, Ra'anana 4353701, Israel}
\affiliation{Astrophysics Research Center of the Open university (ARCO), The Open University of Israel, P.O Box 808, Ra'anana 4353701, Israel}
\affiliation{Department of Physics, The George Washington University, 725 21st Street NW, Washington, DC 20052, USA}

\author[0000-0003-0516-2968]{Ramandeep Gill}
\affiliation{Instituto de Radioastronom\'ia y Astrof\'isica, Universidad Nacional Aut\'onoma de M\'exico, Morelia, Michoac\'an, C.P. 58089, M\'exico}
\affiliation{Astrophysics Research Center of the Open University (ARCO), The Open University of Israel, P.O Box 808, Ra'anana 43537, Israel}

\author[0000-0002-9350-6793]{Tatsuya Matsumoto}
\affiliation{Department of Astronomy, Kyoto University, Kitashirakawa-Oiwake-cho, Sakyo-ku, Kyoto 606-8502, Japan}
\affiliation{Hakubi Center, Kyoto University, Yoshida-honmachi, Sakyo-ku, Kyoto 606-8501, Japan}
\affiliation{Department of Astronomy, School of Science, the University of Tokyo, Bunkyo-ku, Tokyo 113-0033, Japan}

\author[0000-0002-8297-2473]{Kate D. Alexander}
\affiliation{Department of Astronomy and Steward Observatory, University of Arizona, 933 North Cherry Avenue, Tucson, AZ 85721-0065, USA}

\author[0000-0002-0592-4152]{Michael F. Bietenholz}
\affiliation{Department of Physics and Astronomy, York University, 4700 Keele St., Toronto, M3J~1P3, Ontario, Canada}

\author[0000-0003-2349-101X]{Aprajita Hajela}
\affiliation{DARK, Niels Bohr Institute, University of Copenhagen, Jagtvej 155, 2200 Copenhagen, Denmark}

\author[0000-0003-0528-202X]{Collin ~T.~Christy}
\affiliation{Department of Astronomy and Steward Observatory, University of Arizona, 933 North Cherry Avenue, Tucson, AZ 85721-0065, USA}

\author[0000-0002-7706-5668]{Ryan Chornock}
\affiliation{Department of Astronomy, University of California, Berkeley, CA 94720-3411, USA}

\author[0000-0001-6395-6702]{Sebastian Gomez}
\affiliation{Center for Astrophysics | Harvard \& Smithsonian, Cambridge, MA 02138, USA}

\author[0000-0003-0685-3621]{Mark A.~Gurwell}
\affiliation{Center for Astrophysics | Harvard \& Smithsonian, Cambridge, MA 02138, USA}

\author[0000-0002-3490-146X]{Garrett K.~Keating}
\affiliation{Center for Astrophysics | Harvard \& Smithsonian, Cambridge, MA 02138, USA}

\author[0000-0003-1792-2338]{Tanmoy Laskar}
\affiliation{Department of Physics \& Astronomy, University of Utah, Salt Lake City, UT 84112, USA}
\affiliation{Department of Astrophysics/IMAPP, Radboud University, P.O. Box 9010, 6500 GL, Nijmegen, The Netherlands}

\author[0000-0003-4768-7586]{Raffaella Margutti}
\affiliation{Department of Astronomy, University of California, Berkeley, CA 94720-3411, USA}

\author[0000-0002-1407-7944]{Ramprasad Rao}
\affiliation{Center for Astrophysics | Harvard \& Smithsonian, Cambridge, MA 02138, USA}

\author[0009-0005-0038-4141]{Natalie Velez}
\affiliation{Department of Physics, University of Oregon, 1371 E 13th Ave, Eugene OR 97403, USA}

\author[0000-0002-7721-8660]{Mark H. Wieringa}
\affiliation{CSIRO Space and Astronomy,PO Box 76,Epping NSW 1710,Australia}


\begin{abstract}
We present ongoing radio observations of the tidal disruption event (TDE) \tde, which was first detected in the radio at 972 days after disruption, following multiple non-detections from earlier searches.  The new observations presented here span $\approx 1370-2160$ days and $0.88-240$ GHz.  We find that the light curves continue to rise at all frequencies during this time period, following a power law of about $F_\nu\propto t^3$ (compared to $F_\nu\propto t^{5.7}$ at $972-1400$ days), and reaching a peak luminosity of $L\approx 10^{40}$ erg s$^{-1}$, comparable to the luminosity of the relativistic TDE Sw\,1644+57 on the same timescale.  The multi-frequency data indicate that the peak frequency does not significantly evolve over the 1030-day span of our observations, while the peak flux density increases by an order of magnitude.  The observed behavior is consistent with two possible scenarios: (i) a delayed spherical outflow launched about 620 days post-disruption with a velocity of $\approx 0.3c$ and an energy of $\sim10^{50}$ erg, and (ii) a highly off-axis ($\approx 80-90^\circ$) relativistic jet with a Lorentz factor of $\Gamma\sim 8$ and $E_K\approx 10^{52}$ erg. Continued radio observations to capture the light curve peak, as well as VLBI observations, could distinguish between these scenarios.
\end{abstract}


\section{Introduction} 
\label{sec:intro}

A tidal disruption event (TDE) occurs when a star wanders sufficiently close to a supermassive black hole (SMBH) to be torn apart by tidal forces, leading to the eventual formation of a transitory accretion flow \citep{Rees1988,Komossa2015}. Optical/UV and X-ray observations of TDEs are generally thought to track the mass fallback and accretion (e.g. \citealt{Stone2013,Guillochon2013}).  Radio observations, on the other hand, can reveal and characterize the outflows of TDEs in terms of their physical properties \citep{Alexander2020}, including the presence of relativistic jets \citep{Zauderer2011,Giannios11,DeColle2012}.

To date, rapid follow-up of TDEs, within days to weeks after discovery, has led to the radio detection of several events.  These include most prominently the TDE Swift J1644+57 (Sw\,J1644+57), whose radio and mm emission were powered by a relativistic outflow with an energy of $\sim 10^{52}$ erg and an initial Lorentz factor of $\Gamma\sim 10$ \citep{Zauderer2011,Metzger2012,Berger2012,Zauderer2013,Eftekhari2018,Cendes2021,2023MNRAS.524.1386B}. Other events, such as ASASSN-14li and AT2019dsg, have instead exhibited evidence for non-relativistic outflows, with $E_K\sim 10^{48}-10^{49}$ erg and velocity $\beta\approx 0.05-0.1$ \citep[e.g. ][]{Alexander2020,Alexander2016,Cendes2021b,Stein2020}.  In other TDEs, like ASASSN-19bt, the picture is less clear, with an unusual evolution in the radio light curve not fitting either scenario \citep{Christy2024}.

However, in recent years, it is the behavior of TDEs at later times that has proven surprising, with 40\% of all TDEs emitting radio at $\gtrsim$ hundreds of days after optical disruption despite nondetections at earlier epochs \citep{Cendes2024,Horesh2021,Hajela2024}.  Most of these TDEs appear to have non-relativistic outflows, and the mechanism for this emission is not understood, but theories have ranged from changes in density surrounding the SMBH \citep{Matsumoto2024}, a state change in the accretion disc \citep{Hayasaki2021}, the delayed escape of a jet from the vicinity of the black hole  \citep{Teboul2023}, multiple outflows \citep{Goodwin2025}, a combination of origin mechanisms, and others. 

In \citet{p1}, we reported the detection of rapidly rising radio emission from \tde\ ($z=0.0457$) starting about 970 days after optical discovery, with a factor of 30 increase in luminosity compared to upper limits at about 700 days.  Our extensive multi-frequency data led us to the conclusion that the emission was likely due to a mildly relativistic ($\beta\approx 0.2-0.6$) outflow launched with a significant delay of about $700-750$ days after optical discovery, although off-axis jet explanations have also been proposed to explain the emission \citep{Matsumoto2023,Sfaradi2023}.  Further, the late emission from AT2018hyz is inconsistent with the Fundamental Plane- a correlation between radio luminosity, X-ray luminosity, and black hole mass- indicating the unusual nature behind the outflow \citep{Alexander2025}.  Here, we present more recent radio observations of \tde, and evidence that the origin of its outflow- a delayed, mildly relativistic outflow, or a highly off-axis jet launched promptly when the TDE occurred- are both models that fit the data, and the true origin of the outflow will not be clear until we see the end of the rise in the radio light curve.

The paper is structured as follows.  In \S\ref{sec:obs} we describe our new late-time radio, mm, UV/optical, and X-ray observations, and in \S\ref{sec:lumin} we contrast the radio emission from \tde\ with those of previous TDEs.  In \S\ref{sec:modeling} we model the radio spectral energy distribution and carry out an equipartition analysis to derive the physical properties of the outflow and environment for both a spherical outflow and an off-axis jet.  In \S\ref{sec:params} we describe the results for a spherical and a collimated outflow geometry.  We discuss the implications of this outflow in \S\ref{sec:disc} and summarize our findings in \S\ref{sec:conc}.

\section{Observations}
\label{sec:obs}

\subsection{Radio Observations}
\label{sec:obs-radio}

In \citet{p1} we presented observations of \tde\ from $\delta_{t} =972- 1282$ days after optical discovery (on 2018 Oct 14) with the Karl G.~Jansky Very Large Array (VLA), the MeerKAT radio telescope, and the Australian Telescope Compact Array (ATCA).  An error in the VLA calibration pipeline 6.1.2 and 6.2.1\footnote{see: https://science.nrao.edu/facilities/vla/data-processing/pipeline/vla-pipeline-bug-multiband}, discovered after publication of \cite{p1}, resulted in a change in the flux densities at the lowest frequencies ($\approx 1-5$ GHz) for observations at 1126, 1199, and 1251 days.  We include the corrected flux density measurements in Table \ref{tab:obs}.

We obtained multi-frequency observations over 1366-2160 days spanning L- to K-band ($\approx 1-23$ GHz; Programs 22A-458, 22B-205, 23A-241, 23B-056, and 24A-353, PI: Cendes and SC240292, PI: Hajela), which resulted in detections across the full frequency range.  For all observations we used the primary calibrator 3C147, and the secondary calibrator J1024-0052.  We processed the VLA data using standard data reduction procedures in the Common Astronomy Software Application package (CASA; \citealt{McMullin2007}), using {\tt tclean} on the calibrated measurement set available in the NRAO archive, and split the data into subbands by frequency. We obtained all flux densities and uncertainties using the {\tt imtool fitsrc} command within the python-based \texttt{pwkit} package\footnote{https://github.com/pkgw/pwkit} \citep{Williams2017}.  We assumed a point source fit, as preferred by the data.  The observations and resulting flux density measurements are summarized in Table~\ref{tab:obs}.

We obtained observations of \tde\ with the MeerKAT radio telescope in UHF and L-band ($0.8-2$ GHz) on 2022 September 11 (1422 days; DDT-20220414-YC-01, PI: Cendes), and then in UHF band on 2023 January 4 (1543 days; SCI-20220822-YC-01, PI: Cendes) and 2024 February 16 and 2024 September 27 (1951 days and 2175 days, respectively; SCI-20230907-YC-01, PI: Cendes).  For MeerKAT, we used the flux calibrator 0408-6545 and the gain calibrator 3C237, and used the calibrated images obtained via the SARAO Science Data Processor (SDP)\footnote{https://skaafrica.atlassian.net/wiki/spaces/ESDKB/pages/338723406/}.  We confirmed via the secondary SDP products that the source fluxes in the MeerKAT images were consistent with $\sim90\%$ of the sources overlapping with the NRAO VLA Sky Survey \citep[NVSS; ][]{Condon1998}.

We also observed \tde\ with the Australian Telescope Compact Array (ATCA) on 2022 September 1 (1421 days; Program C3472; PI: Cendes) at $2-20$ GHz. For ATCA we reduced the data using the {\tt MIRIAD} package. The calibrator 1934-638 was used to calibrate absolute flux-density and band-pass, while the calibrator 1038+064 was used to correct short term gain and phase changes. The {\tt invert}, {\tt mfclean} and {\tt restor} tasks were used to make deconvolved wideband, natural weighted images in each frequency band. These data are also summarized in Table~\ref{tab:obs}.

\subsection{Millimeter Observations}

We also observed \tde\ with the Atacama Large Millimeter/submillimeter Array (ALMA) at 1367 days and 1441 days (Project 2021.1.01210.T, PI: Alexander). This observation roughly coincides with the multi-frequency VLA observations at 1366 and the ATCA observations at 1421 days. The ALMA observations were in band 3 (mean frequency of 97.5 GHz), and band 6 (240 GHz); the first epoch was in band 3 only.  For the ALMA observations, we used the standard NRAO pipeline (version: 2021.2.0.128) in CASA (version: 6.2.1.7) to calibrate and image the data. 

We also observed \tde\ with the Submillimeter Array (SMA) as part of the POETS program (PI: Berger) at 2198 days, at 225.5 GHz.  The data were processed via the SMA COMPASS pipeline (G. K. Keating et al. 2023, in preparation).

We detect \tde\ in all millimeter observations with a rising flux density over time (Table \ref{tab:obs}).

\subsection{X-ray Observations}

We obtained a new observation of \tde\, at a mean $\delta t = 1934$ days with ACIS-S onboard the \textit{Chandra} X-ray Observatory. The observation was divided into 4 intervals between 2024 February 21 and 2024 February 24 ($\delta t=1933 - 1936$ days),  with respective exposure times of 14.7, 14.9, 14.9, and 6 ks (Program 24500292, PI: Hajela). We reprocessed the observations using the \texttt{repro} task within \texttt{CIAO} 4.15.2 with standard ACIS data filtering and using the latest calibration database (\texttt{CALDB, v4.10.7}). An X-ray source is detected with \texttt{wavdetect} at the position of AT2018hyz in all observations with  high significance ($> 6\sigma$; Gaussian equivalent). The respective 0.5 -- 8 keV net source count-rates in 1.5\arcsec\, region (containing $\gtrsim 90\%$ of the PSF at 1 keV) are: $(1.2\pm0.3)\times 10^{-3}\,\rm{c\,s^{-1}}$, $(0.5\pm0.2)\times 10^{-3}\,\rm{c\,s^{-1}}$, $(0.7\pm0.2)\times 10^{-3}\,\rm{c\,s^{-1}}$, and $(1.0\pm0.4)\times 10^{-3}\,\rm{c\,s^{-1}}$.

We extracted a spectrum of the source with $\texttt{specextract}$ using a 1.5\arcsec\, radius source region and a source-free background region of 30\arcsec\, radius in all four observations. Since there is no statistically significant evidence of variability across the four observations, we fit all the spectra simultaneously with an absorbed simple power-law model ($\texttt{tbabs*ztbabs*pow}$ within $\texttt{Xspec}$). The Galactic neutral hydrogen column density in the direction of AT2018hyz is $N_{H,\rm MW}=2.67\times 10^{20}\,\rm{cm^{-2}}$ \citep{Kalberla05}. Similar to \cite{p1}, we find no evidence for additional intrinsic absorption and thus assume $N_{H, \rm int} = 0\,\rm{cm^{-2}}$. The photon index is constrained to $\Gamma_{\rm X}=1.3_{-0.3}^{+0.5}$ ($1\,\sigma$ confidence level), consistent with the previously reported $\Gamma_{\rm X} = 1.5 \pm 0.7$ in \cite{p1}. The $0.3-10$ keV unabsorbed flux is $F_x=2.0_{-0.3}^{+0.4} \times 10^{-14}\,\rm{erg\,s^{-1}cm^{-2}}$. Assuming a flux density $F_{\rm \nu} \propto \nu^{-\beta_{\rm X}}$, where $\beta_{\rm X} = \Gamma_{\rm X} - 1 = 0.3$, 
 $F_{\rm \nu}$ at $\nu = 1$\,keV is $1.3^{+0.3}_{-0.2}\times10^{-6}$ mJy.

The X-ray flux is consistent with the flux measured at $\delta t=1253$ days \citep{p1}, indicating the X-ray emission may be due to low levels of accretion and unrelated to the TDE itself. We discuss the implications of this in \S\ref{sec:equi}.

\section{Radio Light Curves}
\label{sec:lumin}

\begin{figure}
\begin{center}
    \includegraphics[width=.45\columnwidth]{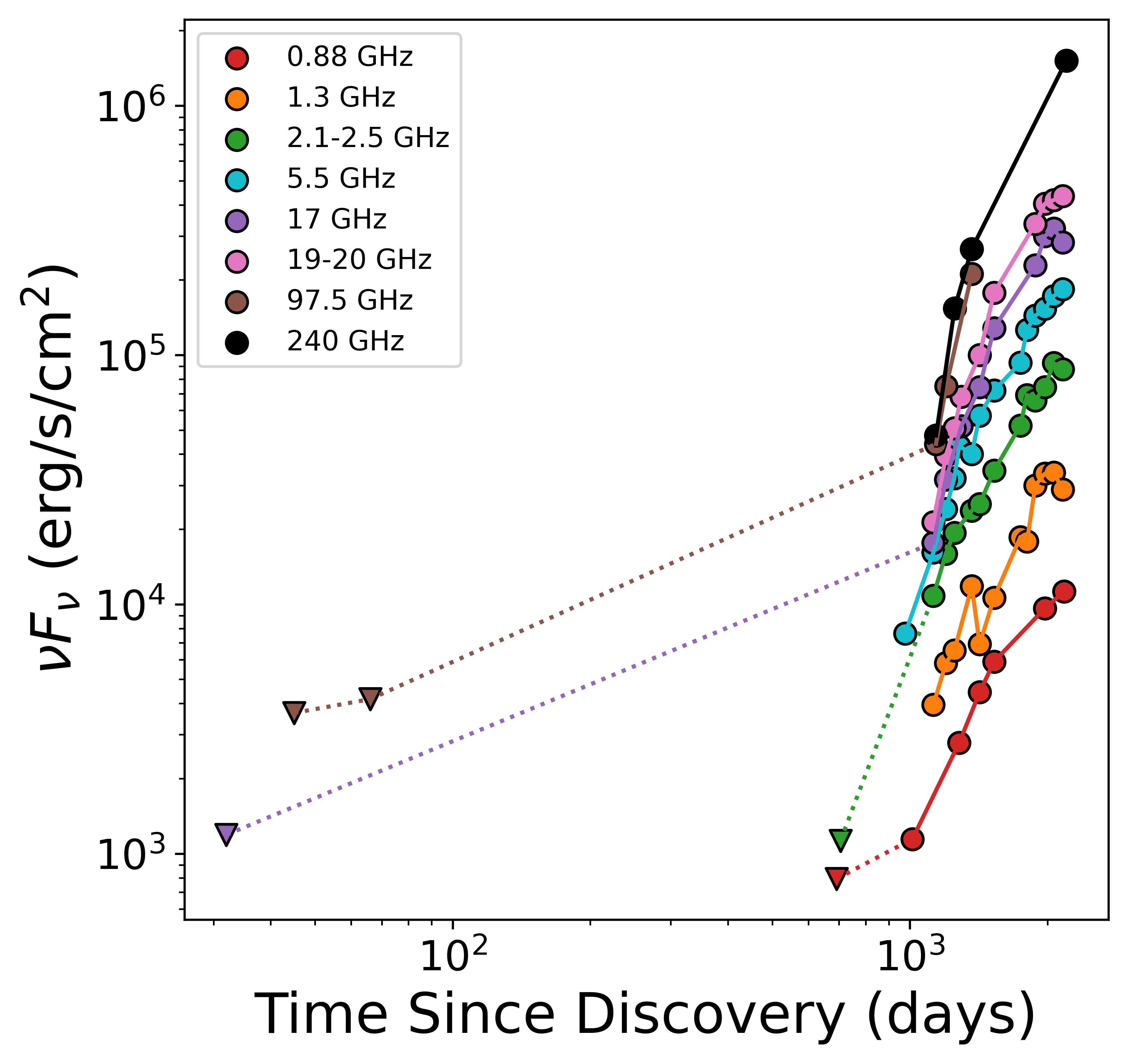}
    \includegraphics[width=.45\columnwidth]{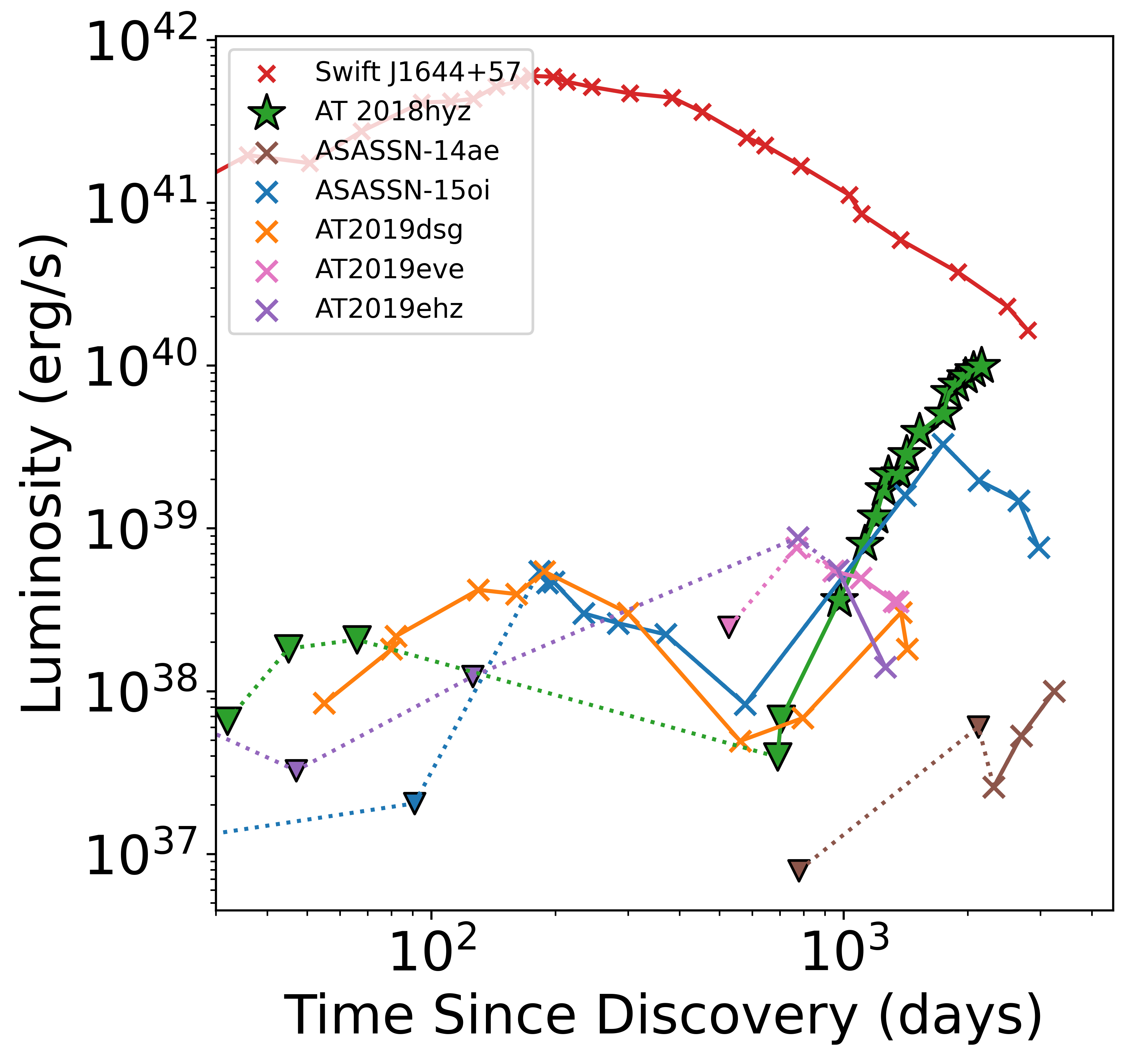}
    \end{center}
    \label{fig:bands}
    \label{fig:lumin-tde}
    \caption{\textit{Left: }Luminosity light curve over time of AT2018hyz in several frequency bands, including early upper limits (triangles) and the late-time detections starting at about 970 days (circles).  While the source is rising in all frequencies since the first radio detections, we find the source has been fluctuating in L-band (1.4 GHz, yellow), and fading in Ku-band (17 GHz, purple) and S-band (3.0 GHz, green) after $\sim2050$ days.  In contrast, at other frequencies such as C-band (5.5 GHz, light blue), K-band (19-20 GHz, pink), UHF-band (0.88 GHz, red) and in the millimeter band (97.5 GHz, brown, and $\sim250$ GHz, black) the source is still rising as roughly $F_\nu\propto t^4$ through 2100 days. \textit{Right: } Luminosity light curve of AT2018hyz, including early upper limits (green triangles; 0.9, 3, and 15 GHz) and the late-time detections starting at about 970 days (green stars; 5 GHz).  Also shown for comparison are the light curves of the relativistic TDE Sw\,J1644+57 (6.7 GHz; red; \citealt{Berger2012,p2,Eftekhari2018,Cendes2021}), the non-relativistic event AT2019dsg (6.7 GHz; orange; \citealt{Cendes2021b,p2}), and four events with apparent late-rising radio emission: ASASSN-15oi (6-7 GHz; blue; \citealt{Horesh2021,Hajela2024}), AT2019eve (5 GHz; pink; \citealt{p2}), AT2019ehz (5 GHz; purple; \citealt{p2}), and ASASSN-14ae (5 GHz; brown; \citealt{p2}).}
\end{figure}

The radio light curves of \tde\ at frequencies of $\approx 0.9-250$ GHz are shown in Figure~\ref{fig:bands}. At all frequencies, we find that the source continues to increase in brightness after the initial detection at 972 days.  At C-band ($5-7$ GHz) we find a steady rise from about 1.4 mJy at 972 days to 33.3 mJy at 2160 days, corresponding to a mean power law rise ($F_\nu \propto t^\alpha$) with $\alpha\approx 4$. We find $\alpha\approx 5.7$ up to about 1400 days, and $\alpha\approx 3.1$ thereafter. We find a similar trend -- rapid initial rise followed by a slightly shallower rise starting at $\approx 1400$ days -- at all well-sampled frequencies.   

In Figure~\ref{fig:lumin-tde} we show the 5 GHz radio light curve in comparison to other radio-emitting TDEs. The radio luminosity of \tde\ increased rapidly by two orders of magnitude from $\lesssim 7\times 10^{37}$ erg s$^{-1}$ at $\approx 700$ days to $\approx 10^{40}$ erg s$^{-1}$ at $\approx 2160$ days. Indeed, it has now exceeded the radio luminosity of any previous TDE other than the relativistic event Sw\,J1644+57, which at a comparable timescale is only about a factor of 3 times more luminous. The rapid rise in \tde\ has continued even beyond the end of a similar rise in ASASSN-15oi at $\approx 500-1400$ days, which has since been steadily fading (Figure~\ref{fig:lumin-tde}; \citealt{Horesh2021,Hajela2024}).  

\begin{figure}
\begin{center}
\includegraphics[width=.3\columnwidth]{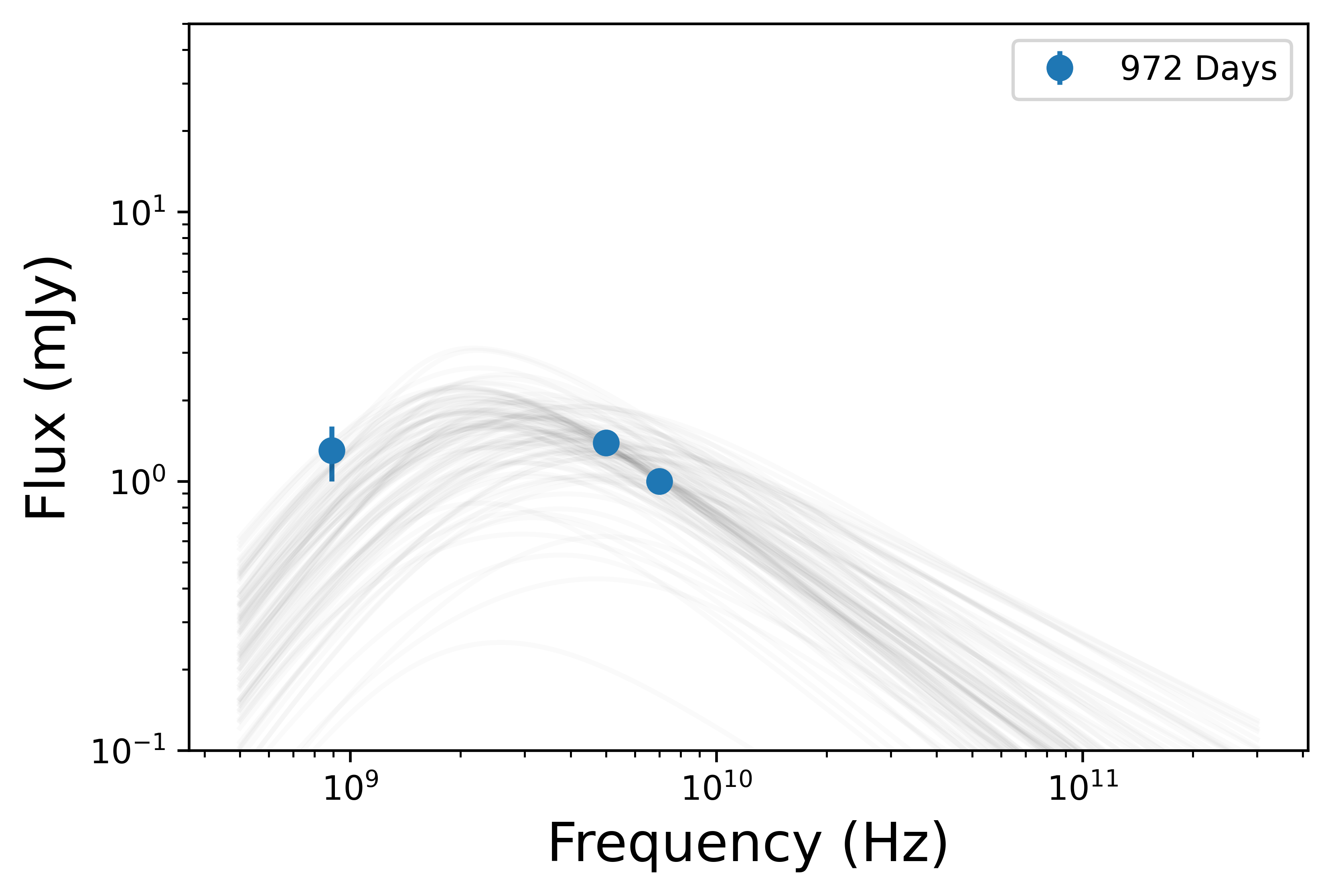}
\includegraphics[width=.3\columnwidth]{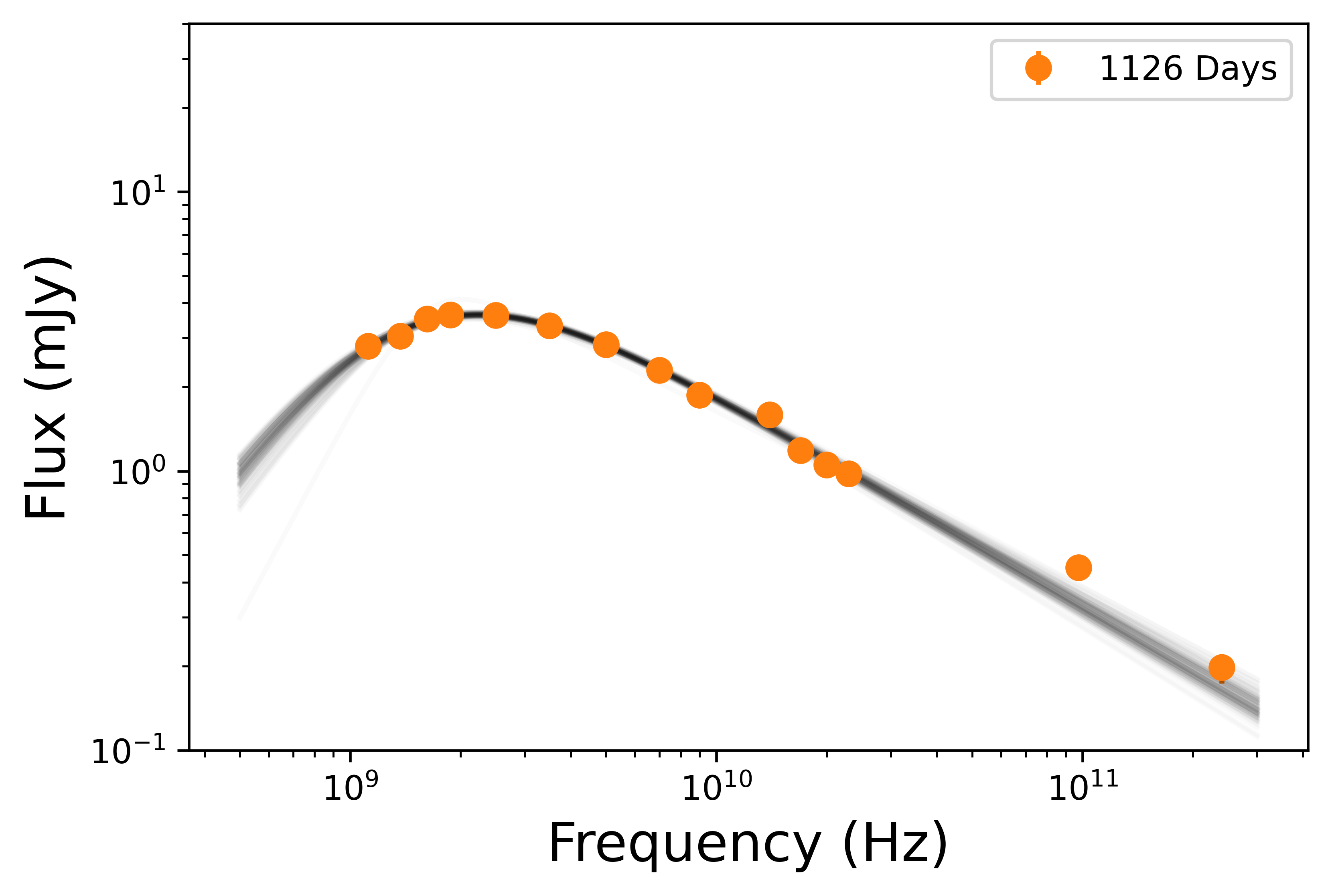}
\includegraphics[width=.3\columnwidth]{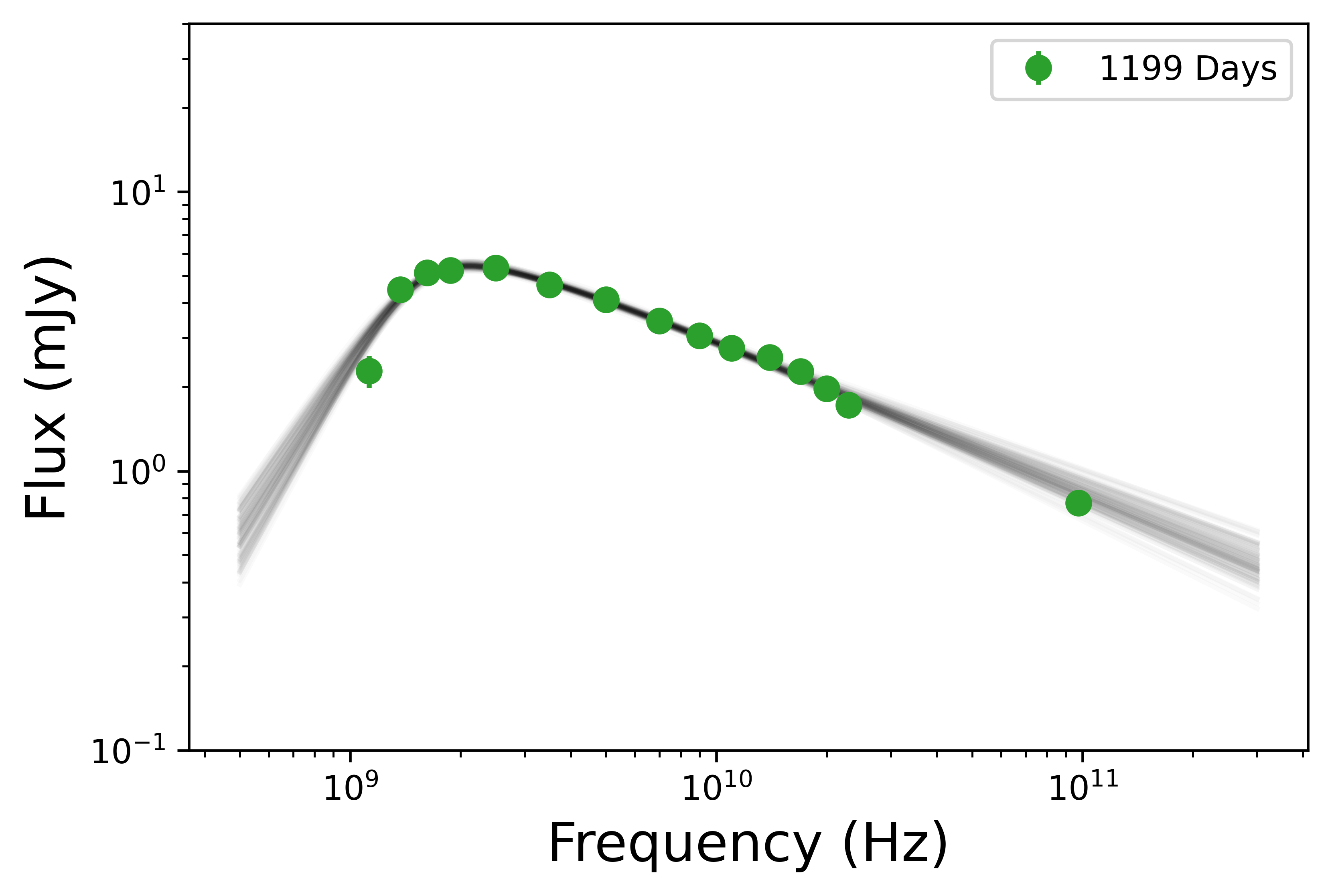}
\includegraphics[width=.3\columnwidth]{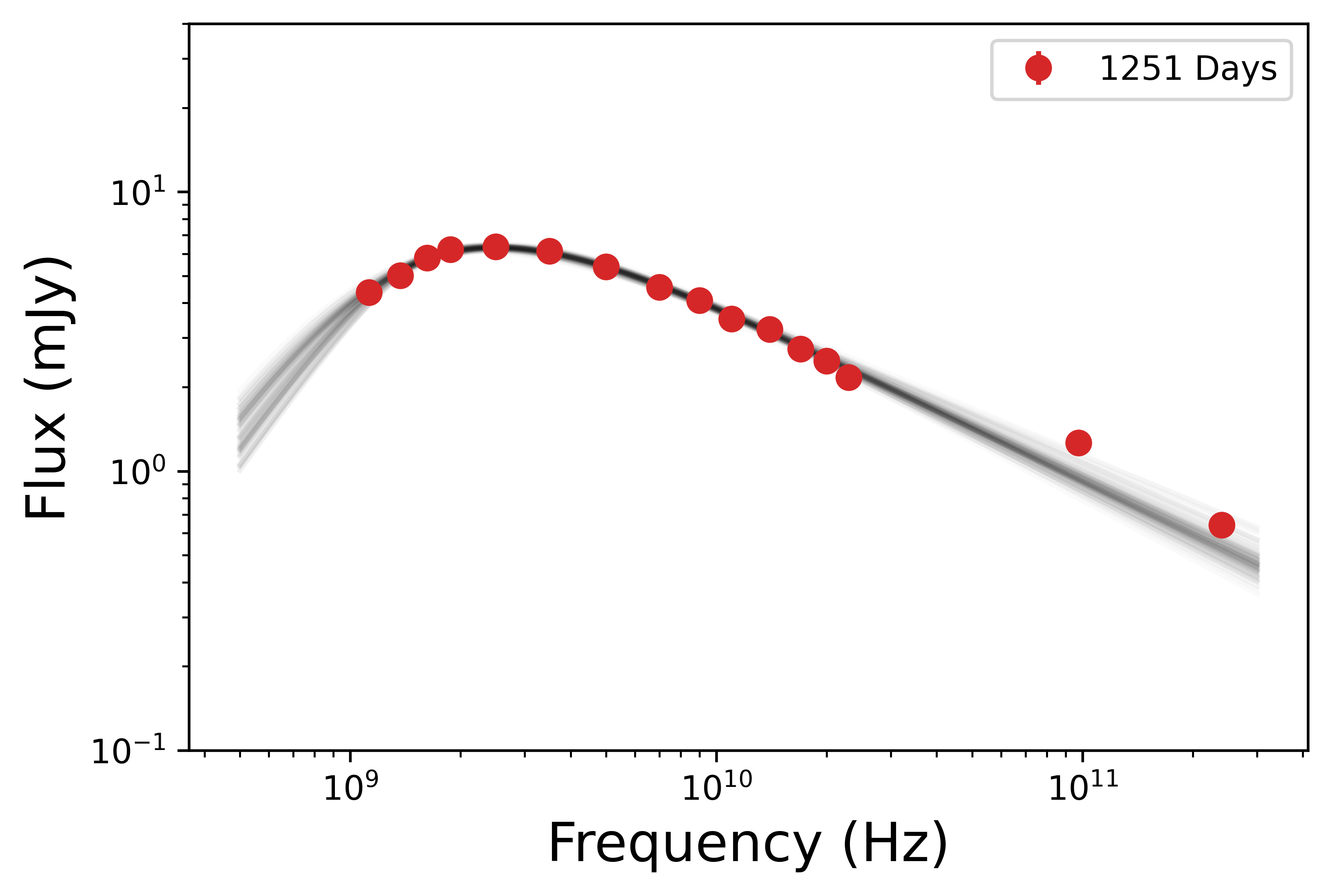}
\includegraphics[width=.3\columnwidth]{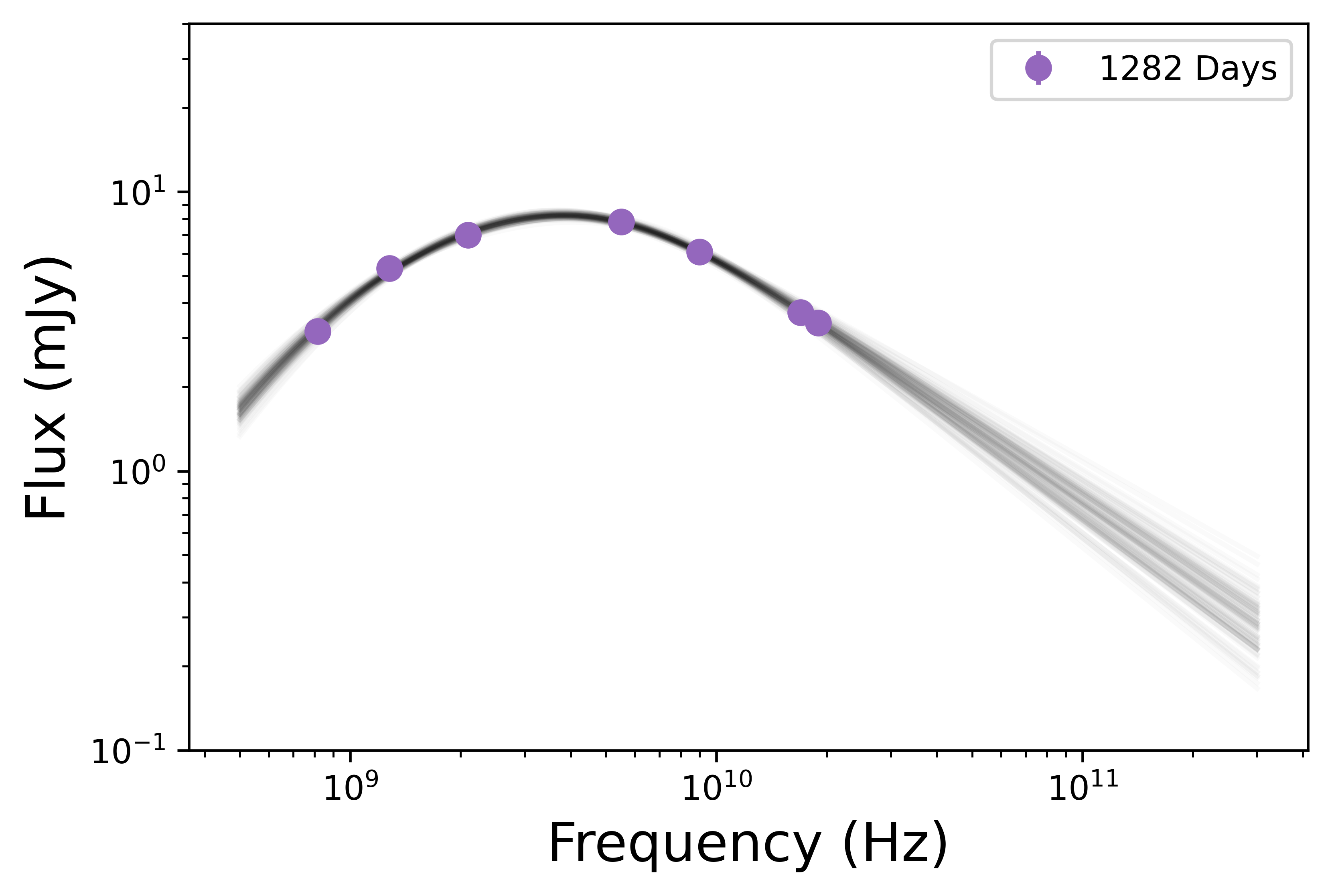}
\includegraphics[width=.3\columnwidth]{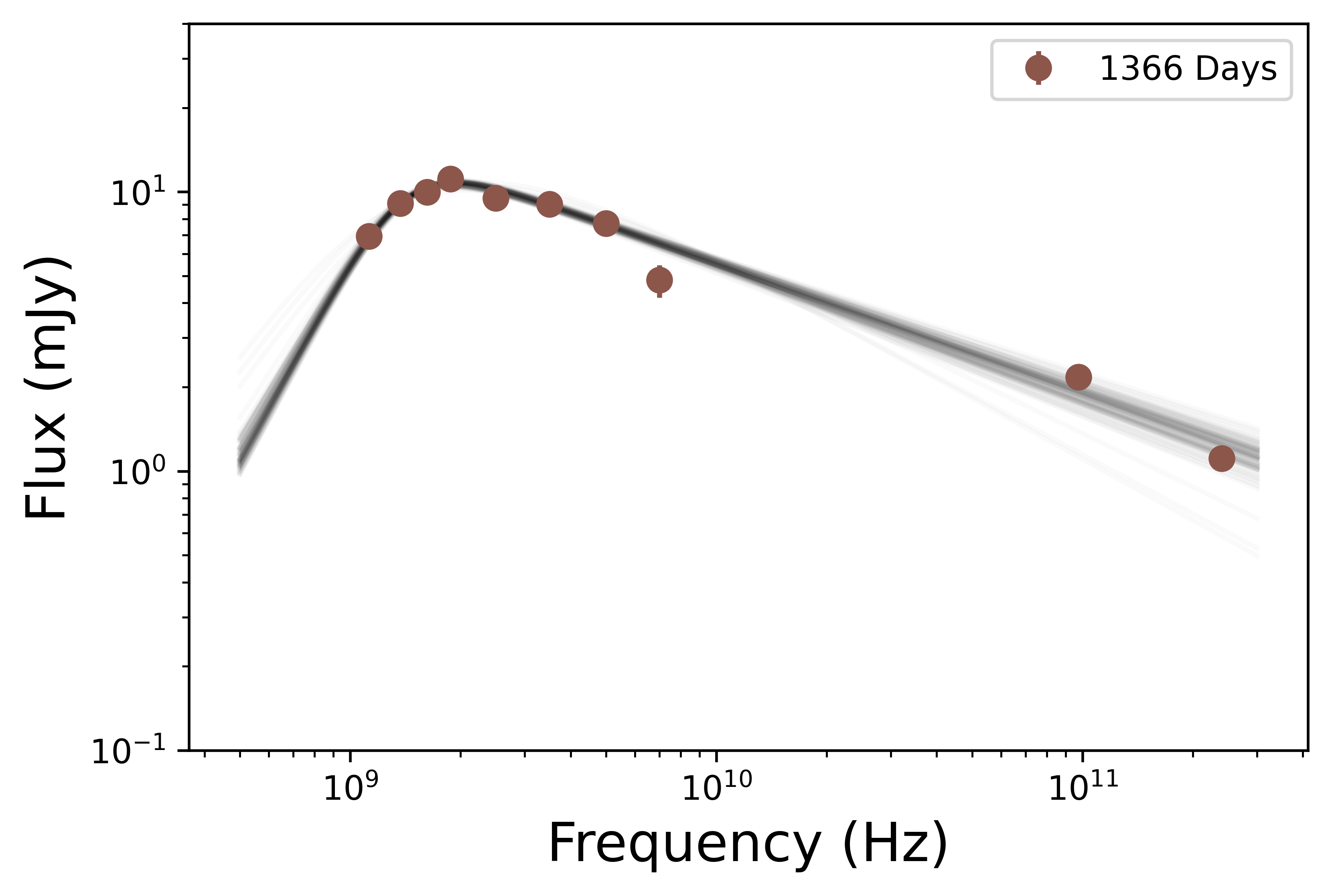}
\includegraphics[width=.3\columnwidth]{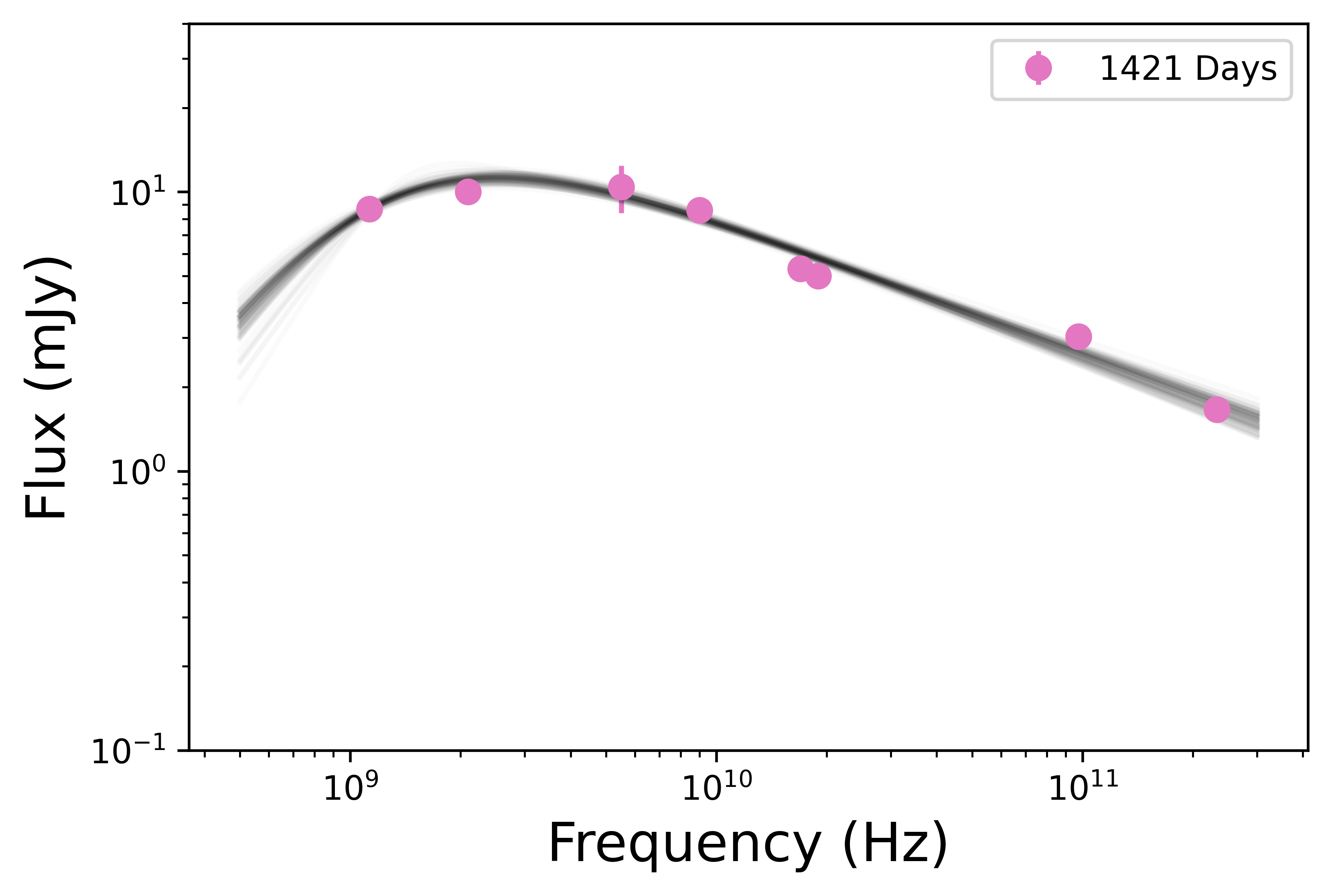}
\includegraphics[width=.3\columnwidth]{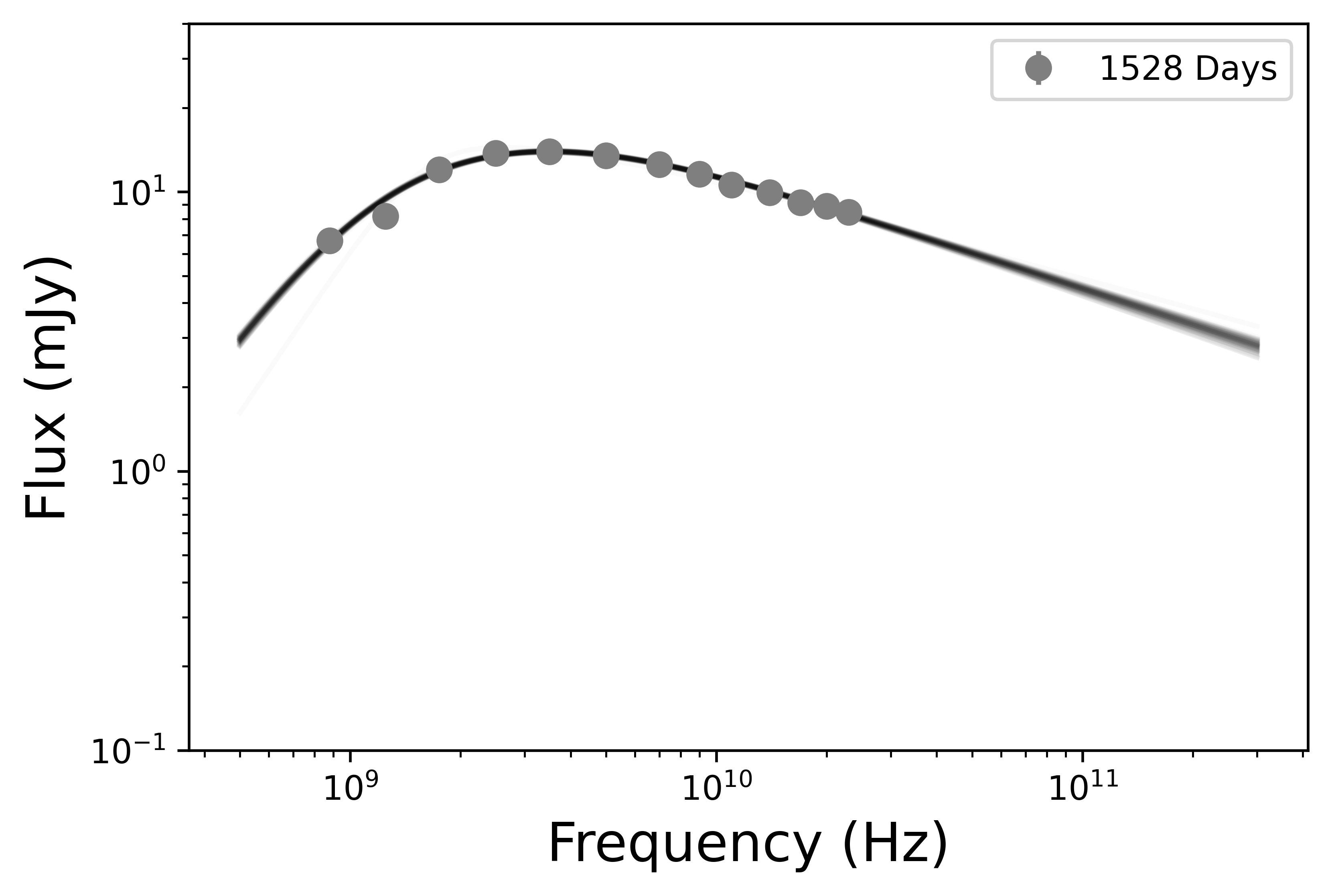}
\includegraphics[width=.3\columnwidth]{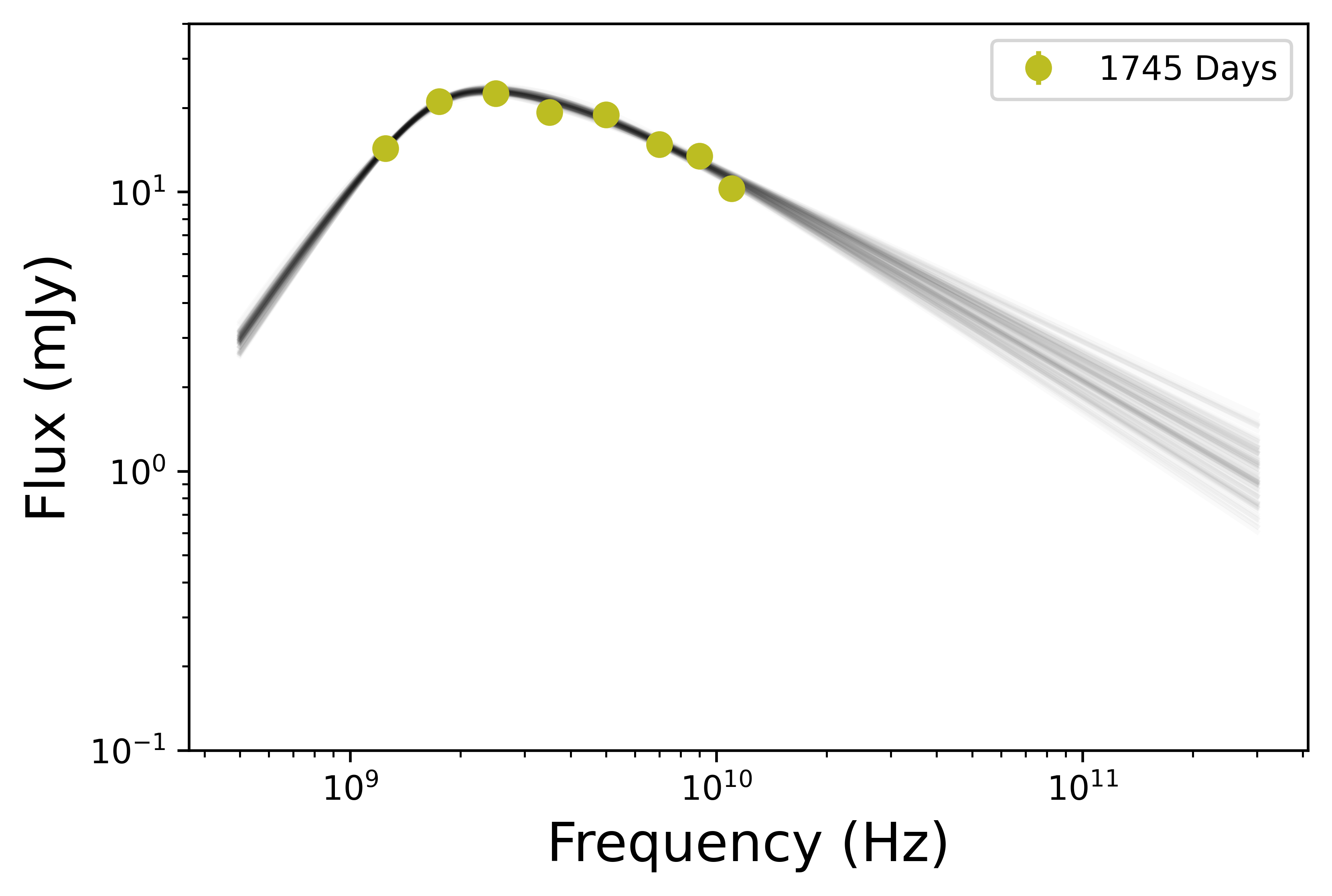}
\includegraphics[width=.3\columnwidth]{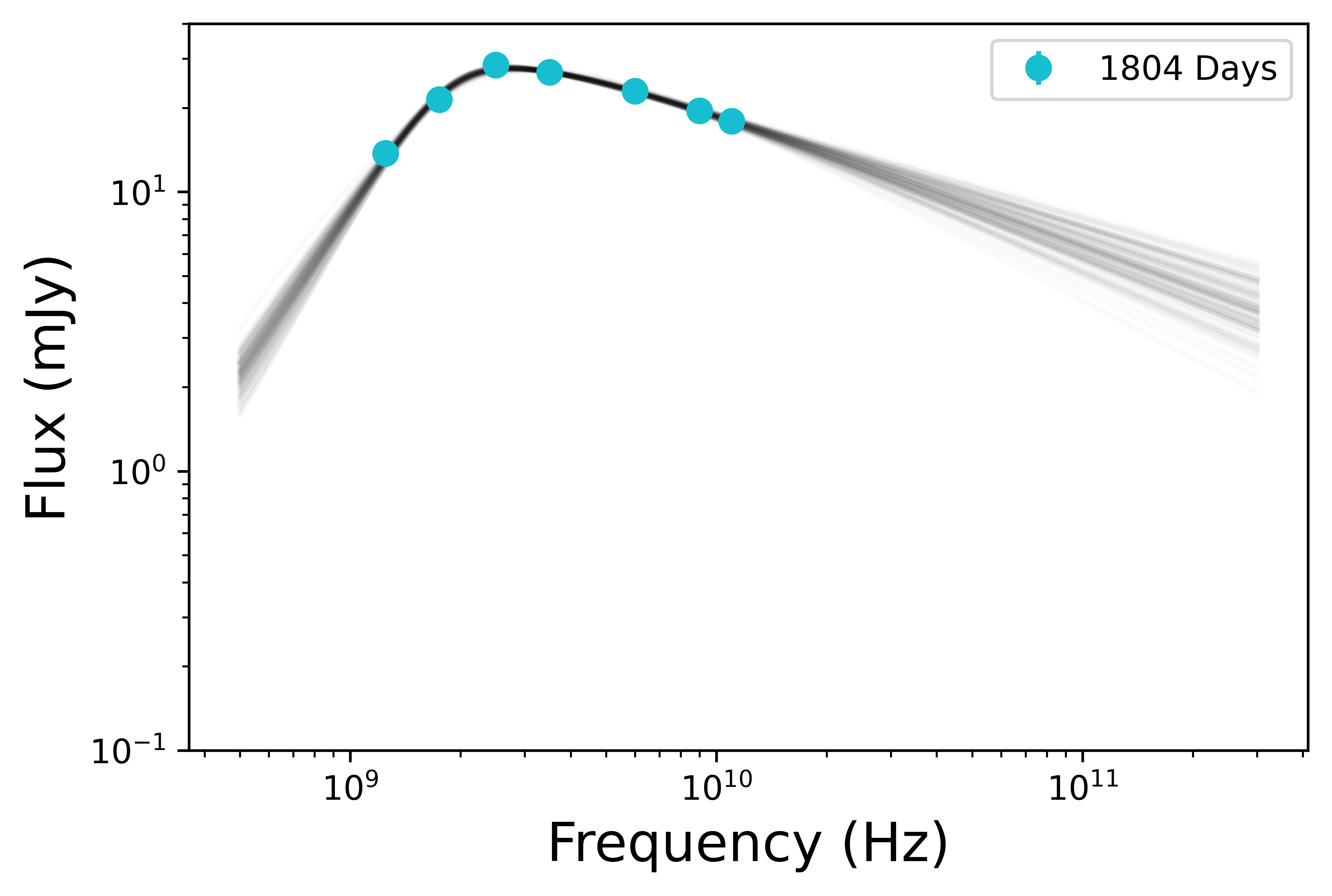}
\includegraphics[width=.3\columnwidth]{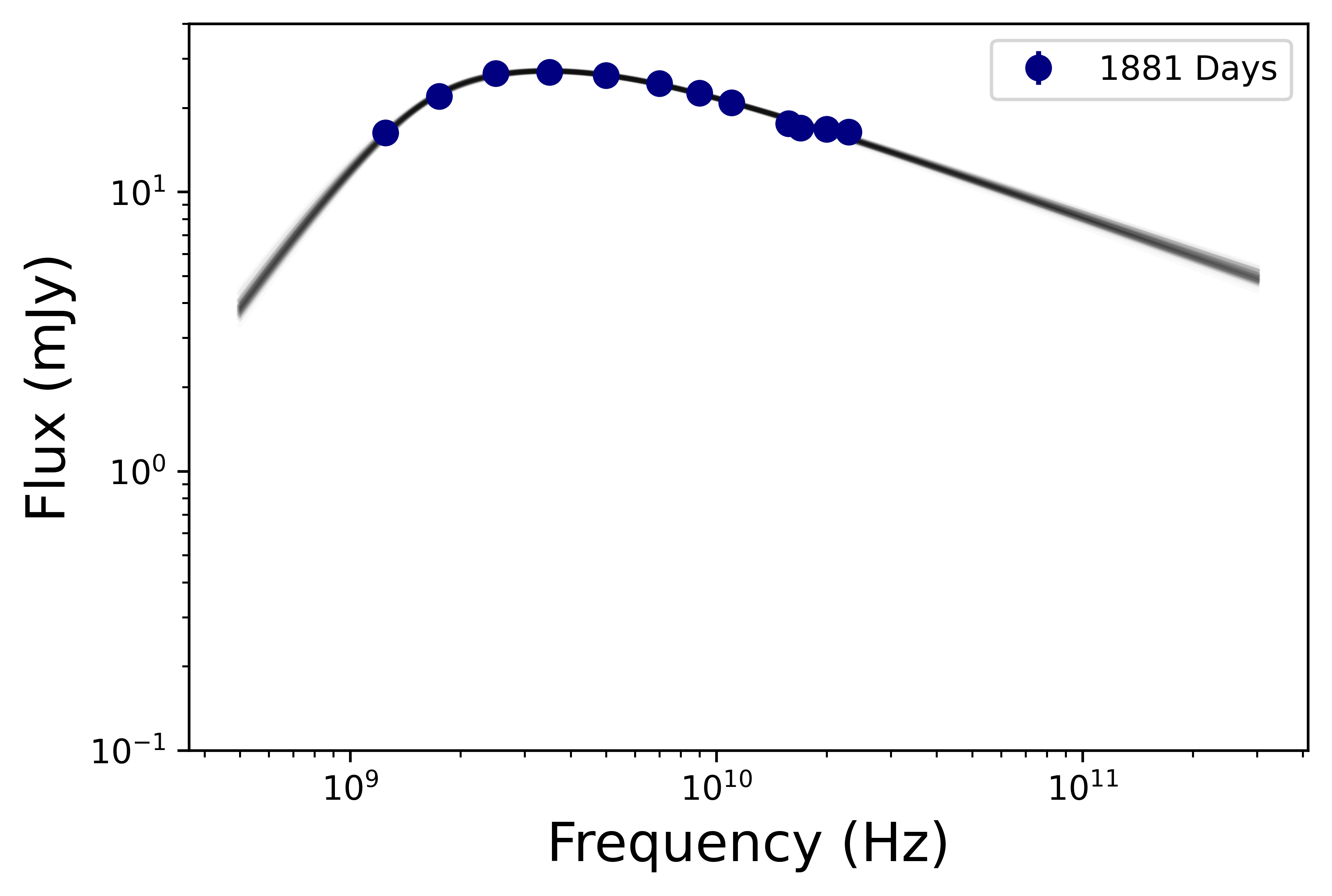}
\includegraphics[width=.3\columnwidth]{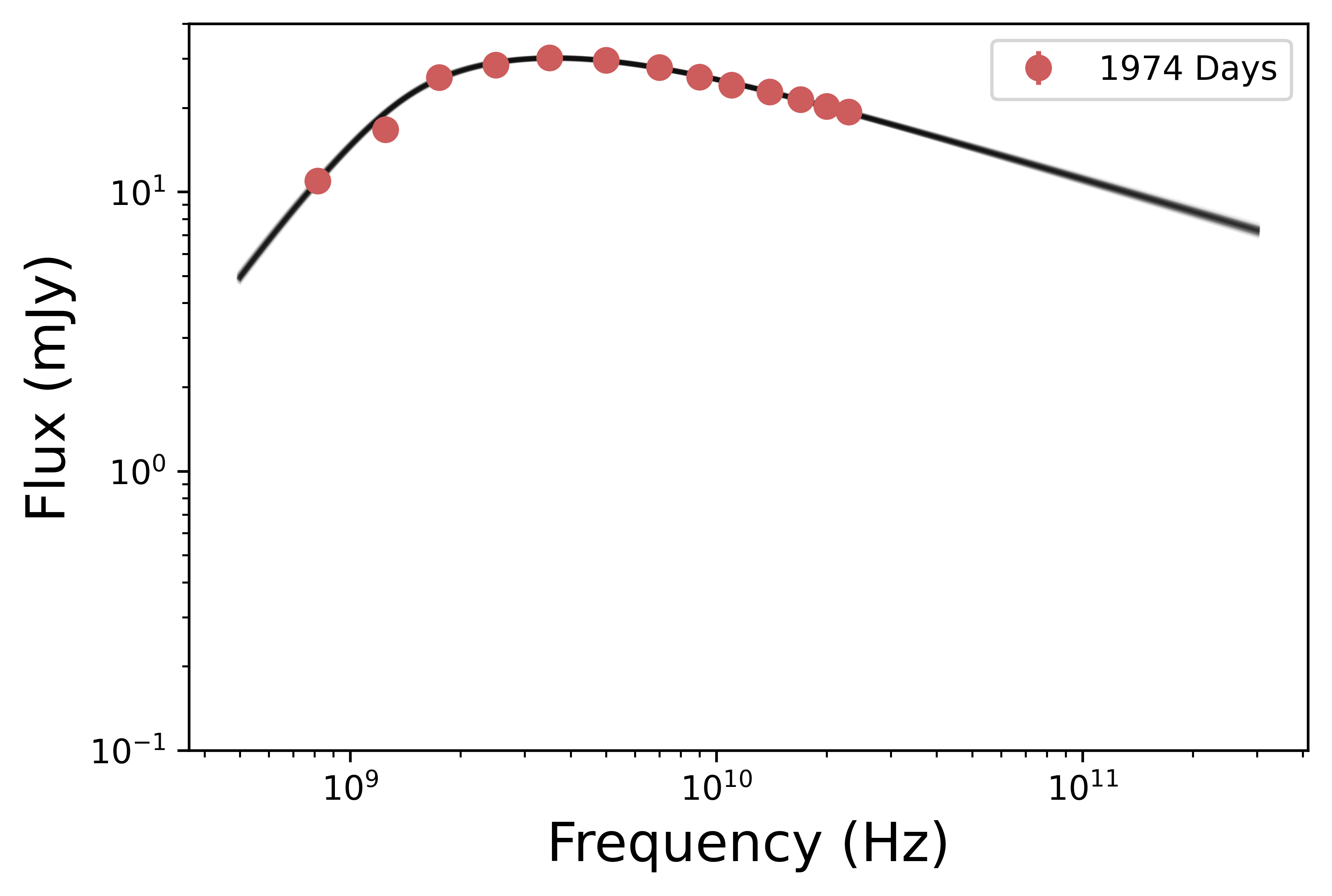}
\includegraphics[width=.3\columnwidth]{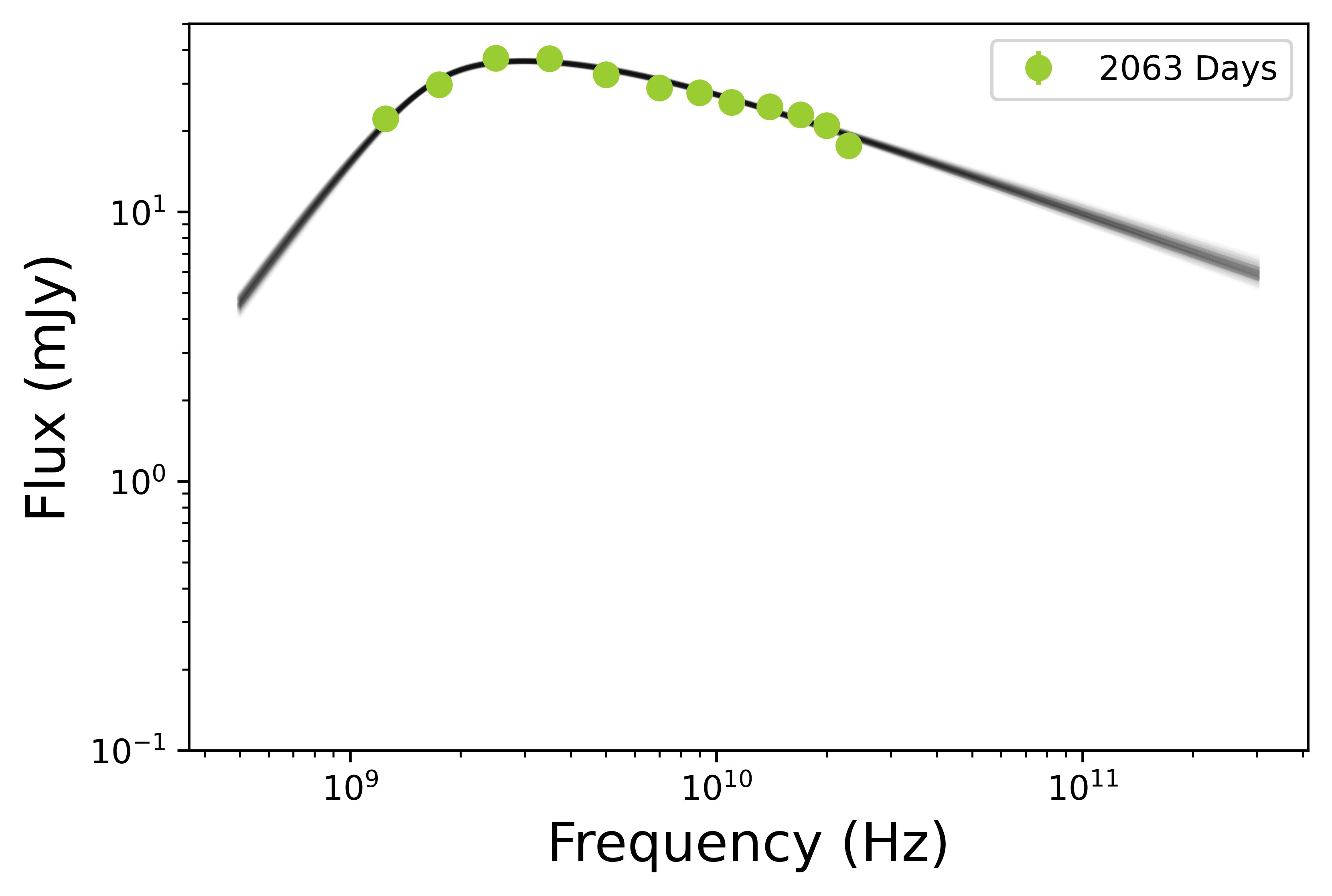}
\includegraphics[width=.3\columnwidth]{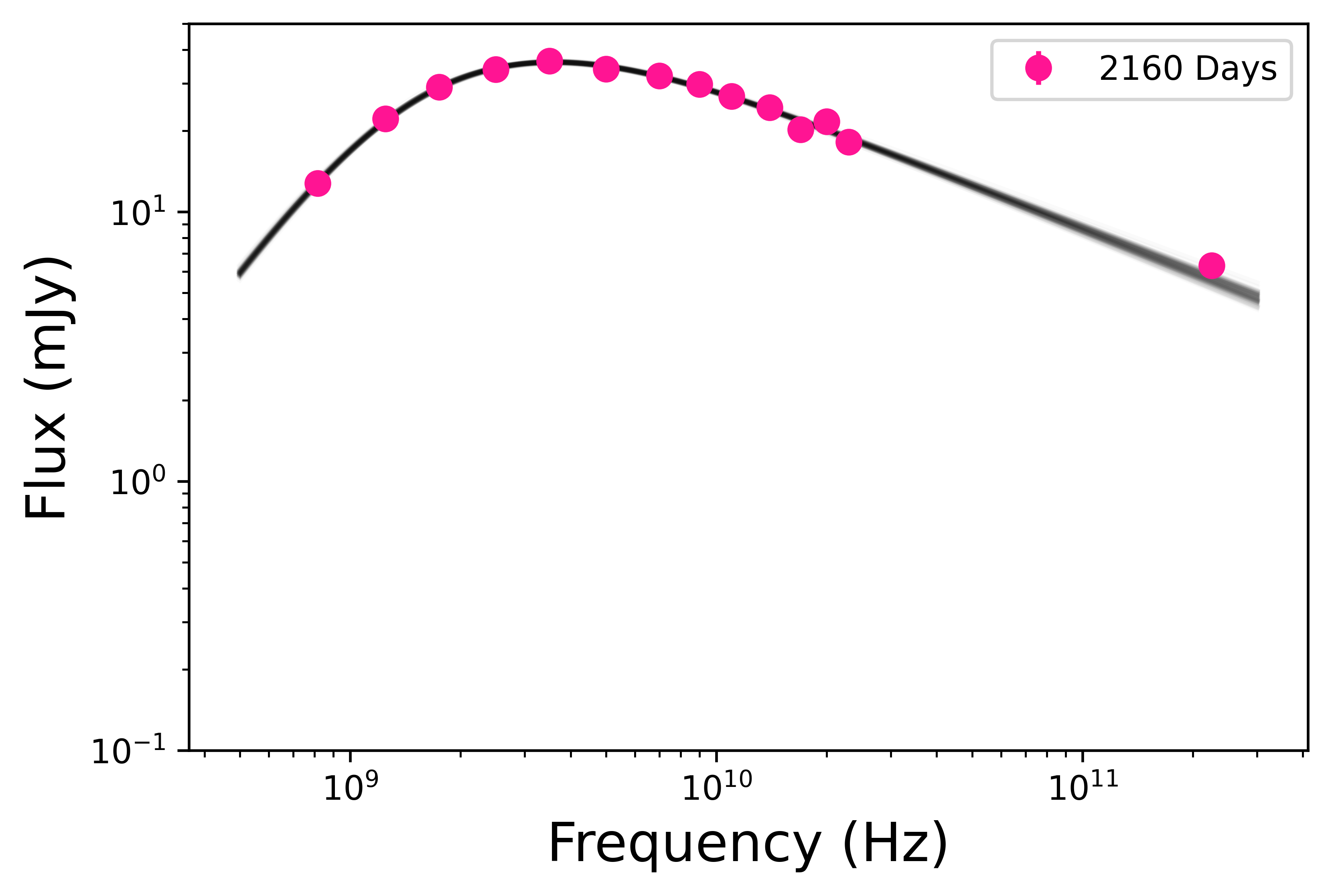}
\end{center}
\label{fig:seds}
\caption{Radio to millimeter spectral energy distributions ranging from 972 days to 2160 days. The grey lines are representative fits from our MCMC modeling (\S\ref{sec:sed}). It is apparent that the epochs exhibit a roughly constant peak frequency ($\approx 1-3$ GHz) and a steadily rising peak flux density.  SED light curves are combined with observations occurring within a few weeks of other observations as outlined in Table \ref{tab:obs} and \citet{p1}.}
\end{figure}

\begin{deluxetable*}{lcccc}
\label{tab:sed}
\tablecolumns{5}
\tablecaption{Spectral Energy Distribution Parameters}
\tablehead{
	\colhead{$\delta t$} &
	\colhead{$F_{\nu,p}$} &
	\colhead{$\nu_p$} &
	\colhead{$p$} &
	\colhead{$\sigma$} 
	\\
    \colhead{(d)} &
    \colhead{(mJy)} &
    \colhead{GHz} &
    \colhead{} &
    \colhead{} 
}
\startdata
972\,$^a$ & $2.38\pm 0.35$ & $1.62\pm 0.05$ & 2.3 & $0.19\pm 0.17$ \\
1126 & $3.70\pm 0.05$ & $2.32\pm 0.05$ & $2.52\pm 0.05$ & $0.06\pm 0.02$\\
1199 & $5.26\pm 0.16$ & $2.10\pm 0.08$ & $2.09\pm 0.08$ & $0.08\pm 0.03$\\
1251 & $6.45\pm 0.10$ & $2.63\pm 0.07$ & $2.27\pm 0.07$ & $0.14\pm 0.03$\\
1282 & $8.83\pm 0.68$ & $3.41\pm 0.04$ & $2.58\pm 0.21$ & $0.14\pm 0.04$\\
1366 & $11.10\pm 0.23$ & $1.86\pm 0.13$ & $1.94\pm 0.13$ & $0.16\pm 0.04$\\
1421 & $10.12\pm 0.24$ & $2.30\pm 0.14$ & $1.99\pm 0.14$ & $0.09\pm 0.02$\\
1528 & $14.04\pm 0.39$ & $3.57\pm 0.06$ & $1.89\pm 0.06$ & $0.04\pm 0.02$\\
1745 & $23.42\pm 2.58$ & $2.41\pm 0.20$ & $2.48\pm 0.20$ & $0.06\pm 0.02$\\
1804 & $28.16\pm 1.37$ & $2.71\pm 0.05$ & $2.02\pm 0.13$ & $0.07\pm 0.03$\\
1881 & $26.93\pm 0.77$ & $3.37\pm 0.03$ & $1.92\pm 0.05$ & $0.05\pm 0.03$\\
1974 & $30.03\pm 0.37$ & $3.61\pm 0.04$ & $1.77\pm 0.03$ & $0.14\pm 0.03$\\
2063 & $33.23\pm 0.68$ & $3.21\pm 0.02$ & $1.88\pm 0.07$ & $0.16\pm 0.05$\\
2160 & $36.01\pm 1.29$ & $3.56\pm 0.04$ & $2.08\pm 0.03$ & $0.07\pm 0.04$\\
\enddata
\tablecomments{$^a$ The parameters for this observation were not affected by the VLA calibration pipeline error outlined in Section \ref{sec:obs-radio} and is the same as reported in \citet{p1}; we include it for completion.}
\end{deluxetable*}

\section{Modeling and Analysis}
\label{sec:modeling}

\subsection{Spectral Energy Distributions}
\label{sec:sed}

Following our previous work \citep{Eftekhari2018,Cendes2021,Cendes2021b,p1,p2}, we fit the spectral energy distributions (SEDs) with a weighted model that accounts for two possible origins for the peak frequency --- $\nu_m$ (i.e., relativistic outflow) or $\nu_a$ (i.e., non-relativistic outflow) --- using the models developed by \citet{Granot2002} for synchrotron emission from GRB afterglows.  When $\nu_{a}\ll\nu_{m}$ the spectrum is given by:
\begin{equation}
F_1 \equiv F_\nu (\nu_a) \Bigg[ \Big(\frac{\nu}{\nu_a}\Big)^{-2s_1}+\Big(\frac{\nu}{\nu_a}\Big)^{- s_1/3}\Bigg] \ \times 
\Bigg[1+\Big(\frac{\nu}{\nu_m}\Big)^{s_2 (\frac{1}{3}-(1-p)/2)}\Bigg]^{-1/s_2},
\label{eq:first-spec}
\end{equation}
where $s_1= 1.06$ and $s_2= 1.76 + 0.05p$ are smoothing parameters that describe the shape of the spectrum across each break. When $\nu_{m}\ll\nu_{a}$, the spectrum is instead given by:
\begin{equation}
F_2 \equiv F_\nu (\nu_m) \Bigg[ \Big(\frac{\nu}{\nu_m}\Big)^2 e^{-s_4 (\frac{\nu}{\nu_m})^{2/3}}+\Big(\frac{\nu}{\nu_m}\Big)^{5/2}\Bigg] \ 
\times
\Bigg[1+\Big(\frac{\nu}{\nu_a}\Big)^{s_2 (2.5-(1-p)/2)}\Bigg]^{-1/s_2},
\label{eq:second-spec}
\end{equation}
Where $s_4 = 3.63p - 1.60$.  To obtain a smooth transition between the two regimes, we employ the following weighting \citep{Eftekhari2018}:
\begin{equation}
F = \frac{w_1F_1 + w_2F_2}{w_1 + w_2},
\label{eq:weighted}
\end{equation}
where $w_1 = (\nu_m/\nu_a)^2$ and $w_2 = (\nu_a/\nu_m)^2$. 

We determine the best fit parameters of the model --- $F_\nu(\nu_m)$, $\nu_a$, $\nu_m$ and $p$ --- using the Python Markov Chain Monte Carlo (MCMC) module \texttt{emcee} \citep{Foreman-Mackey2013}, assuming a Gaussian likelihood where the data have a Gaussian distribution for the parameters $F_{\nu}(\nu_m)$ and $\nu_a$, as in \citet{Cendes2021}.  For $p$ we use a uniform prior of $p=1.5-3.5$. We also include in the model a parameter that accounts for additional systematic uncertainty beyond the statistical uncertainty on the individual data points, $\sigma$, as a fractional error added to each data point.  The posterior distributions are sampled using 100 MCMC chains, which were run for 2,000 steps, discarding the first 1,000 steps to ensure the samples have sufficiently converged by examining the sampler distribution. The resulting SED fits are shown in Figure~\ref{fig:seds} and provide a good fit to the data. 

\begin{figure}[t!]
\begin{center}
\includegraphics[width=.32\columnwidth]{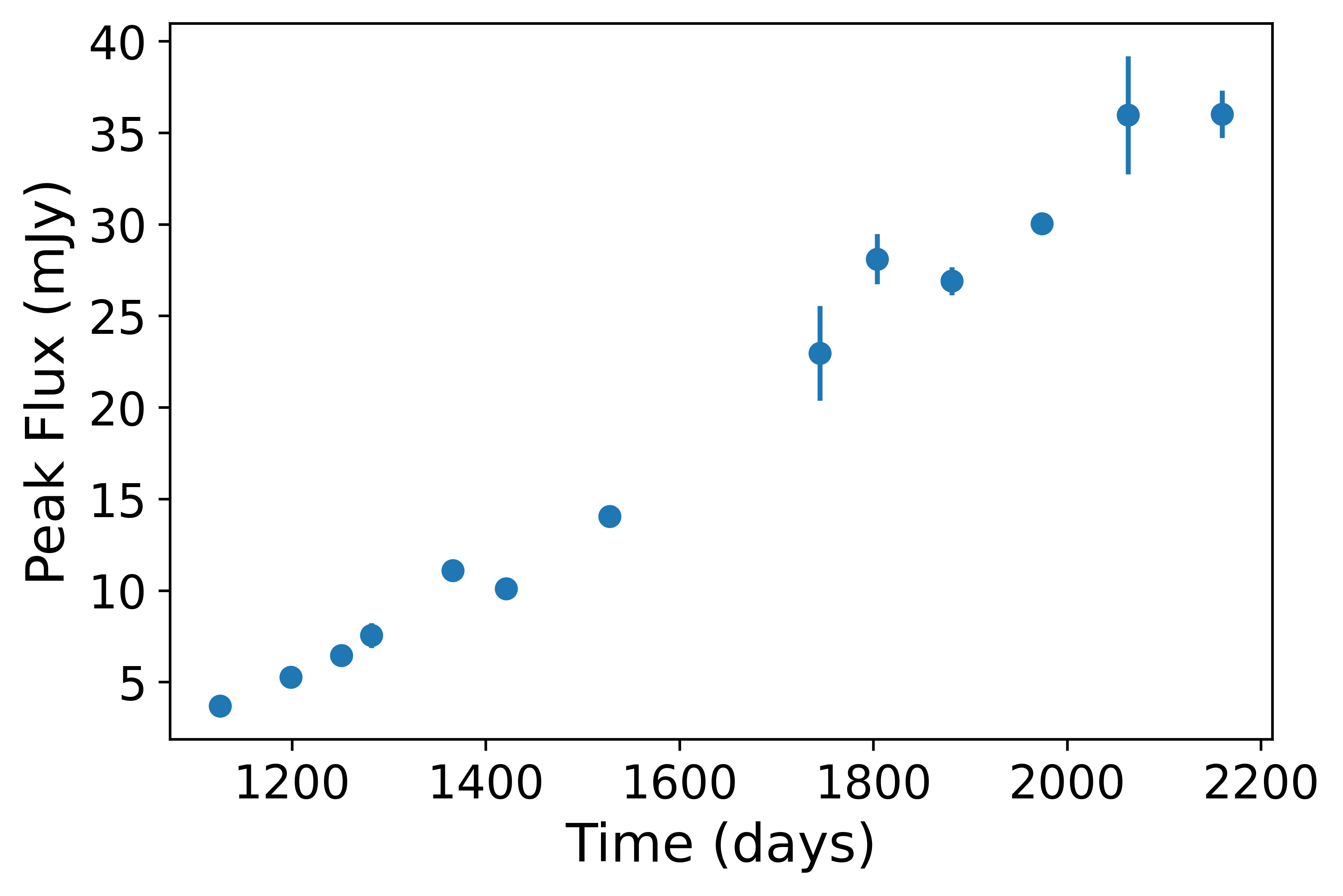}
\includegraphics[width=.32\columnwidth]{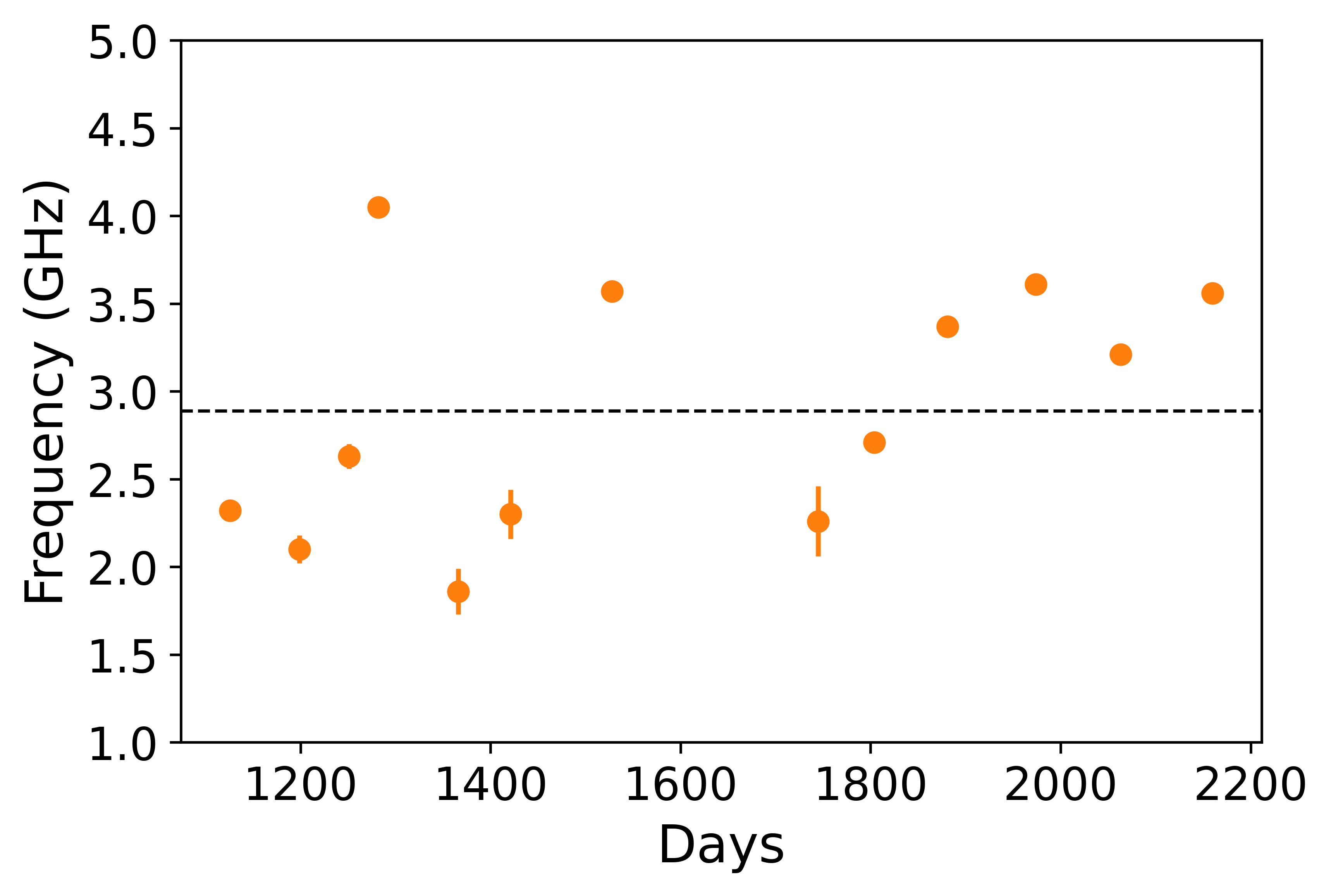}
\includegraphics[width=.32\columnwidth]{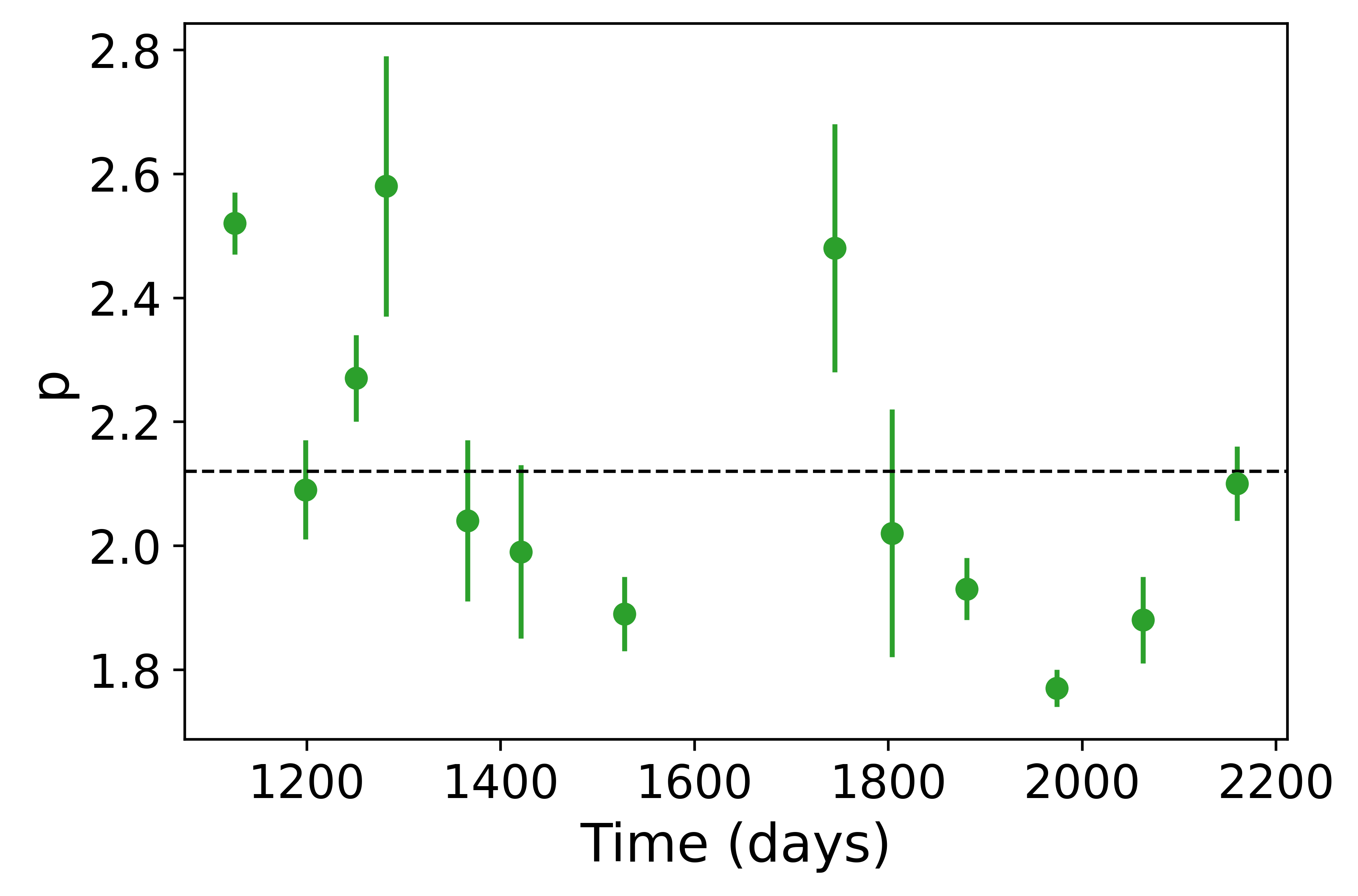}
\end{center}
\label{fig:params}
\caption{The evolution in peak flux ($F_p$; blue), peak frequency ($\nu_p$; orange) and $p$ (green) over time for \tde.  We find that $F_p$ steadily increases over time, whereas $\nu_p$ and p have variations over time dominated by epoch-to-epoch scatter.  We include a dashed line at the mean value for these parameters $\nu_p = 2.9$ GHz and $p= 2.16$.}
\end{figure}

Our SED modeling indicates that \tde\ is in the regime where $\nu_a \approx \nu_m$, with $\nu_a\approx$ 1.5 GHz and $\nu_m \approx$ 3 GHz.  While we fit the data at 972 days, we exclude it from all subsequent averages due to a paucity of data and thus higher error compared to the other observations.  From the SED fits we determine the peak frequency and flux density, $\nu_p$ and $F_{\nu,p}$, respectively, which are used as input parameters for the determination of the physical properties (\S\ref{sec:equi}).  The best-fit values and associated uncertainties are listed in Table~\ref{tab:sed} and are shown in Figure~\ref{fig:params}.  While $F_{\nu,p}$ exhibits a steady rise over time (by a factor of $\approx10$), $\nu_p$ remains steady overall (despite some epoch-to-epoch scatter) with a mean value of $2.9\pm 0.7$ GHz (Figure \ref{fig:params}). The value of $p$ exhibits some scatter as well, but remains steady, with a mean of $2.16\pm 0.26$; we thus adopt $p=2.16$ in the subsequent analysis.

\subsection{Spherical Outflow Equipartition Analysis}
\label{sec:equi}

Using the inferred values of $\nu_p$, $F_{\nu,p}$ and $p$, we derive the physical properties of the outflow and ambient medium using a minimum energy analysis assuming a spherical geometry, following the method outlined in \citet{p1} (with the exception of an updated value for the fraction of post-shock energy in magnetic fields of $\epsilon_{B}=10^{-3}$; see below).  Given the steady value of $\nu_p$ (with variations dominated by epoch-to-epoch scatter), we fix $\nu_p=2.9$ GHz to focus on the long-term trends in the physical parameters (Figure~\ref{fig:params}).  We also add the RMS for$\nu_p$, 0.6 GHz, to the error when inferring the physical parameter uncertainties to account for the error due to its variation over time.  We infer the radius ($R$), kinetic energy ($E_K$), the magnetic field strength ($B$), the number of radiating electrons ($N_e$), and the Lorentz factor ($\Gamma$) of electrons radiating at $\nu_a$ ($\gamma_a$) \citep{Duran2013}. The resulting parameters are listed in Table~\ref{tab:equi}.

Using the inferred physical parameters, we can also predict the location of the relativistic synchrotron cooling frequency as outlined in \citet{p1}, and given by \citep{Sari1998}:
\begin{equation} 
\nu_c\approx 2.25\times 10^{14} B^{-3} \Gamma^{-1} t_d^{-2}\,\,{\rm Hz}.
\label{eq:cooling}
\end{equation}
Since $\nu_c$ has a strong dependence on the magnetic field strength, measuring its location directly from the data can help to determine the value of $\epsilon_B$ and whether the outflow deviates from equipartition. Our \textit{Chandra} observation at 1934 days allows us to complete a full SED when combined with radio data at 1974 days.  We find that as in \cite{p1}, a model that does not include a cooling break over-predicts the \textit{Chandra} measurement.  The required steepening is indicative of a cooling break, which allows us to fit an additional multiplicative term as outlined in \citet{p1}. Fitting this model to the data we find $\nu_c\approx 250$ GHz at $\approx 1950$ days.  This is consistent with our VLA+SMA observations at 2160 days, which do not exhibit a cooling break between the radio and mm bands. With the value of $\nu_c$ determined, we find that $\epsilon_{B}\approx 10^{-3}$. This is similar to the value inferred for Sw\,J1644+57 \citep{Eftekhari2018,Cendes2021}.  The values listed in Table~\ref{tab:equi} are adjusted for this value of $\epsilon_B$.

\begin{deluxetable*}{llllllllll}
\label{tab:equi}
\tablecolumns{10}
\tablecaption{Equipartition Model Parameters}
\tablehead{
	\colhead{$\delta t$} &
	\colhead{log($R_{\rm eq}$)} &
	\colhead{log($E_{\rm eq}$)} &
	\colhead{log($B$)} &
	\colhead{log($N_e$)} &
	\colhead{log($n_{\rm ext}$)} &
	\colhead{$\Gamma$} &
	\colhead{$\beta$} &
	\colhead{$\gamma_a$} &
	\colhead{$\nu_c$} 
	\\
    \colhead{(d)} &
    \colhead{(cm)} &
    \colhead{(erg)} &
    \colhead{(G)} &
    \colhead{} &
    \colhead{(cm$^{-3}$)} &
    \colhead{} &
    \colhead{} &
    \colhead{} &
    \colhead{(GHz)} 
}
\startdata
 972 &$17.50\substack{+0.13 \\ -0.11}$&$48.74\substack{+0.21 \\ -0.15}$&$-1.05\substack{+0.01 \\ -0.01}$&$53.49\substack{+0.15 \\ -0.09}$&$0.40\substack{+0.01 \\ -0.03}$&$1.04\substack{+0.01 \\ -0.01}$&$0.33\substack{+0.02 \\ -0.01}$ & $108\substack{+2 \\ -1} $& $2504\substack{+212 \\ -122}$ \\
1126 &$17.66\substack{+0.09 \\ -0.10}$&$49.14\substack{+0.09 \\ -0.09}$&$-1.08\substack{+0.01 \\ -0.01}$&$53.89\substack{+0.03 \\ -0.03}$&$0.32\substack{+0.01 \\ -0.01}$&$1.04\substack{+0.01 \\ -0.01}$&$0.33\substack{+0.01 \\ -0.01}$& $113 \substack{+1 \\ -1}$ &$1583\substack{+14 \\ -13}$ \\
1199 &$17.74\substack{+0.09 \\ -0.09}$&$49.33\substack{+0.09 \\ -0.09}$&$-1.11\substack{+0.01 \\ -0.01}$&$54.09\substack{+0.03 \\ -0.03}$&$0.29\substack{+0.01 \\ -0.01}$&$1.04\substack{+0.01 \\ -0.01}$&$0.34\substack{+0.01 \\ -0.01}$& $115 \substack{+1 \\ -1}$ &$1405\substack{+13 \\ -13}$\\
1251 &$17.78\substack{+0.09 \\ -0.10}$&$49.45\substack{+0.09 \\ -0.09}$&$-1.12\substack{+0.01 \\ -0.01}$&$54.20\substack{+0.03 \\ -0.03}$&$0.27\substack{+0.01 \\ -0.01}$&$1.04\substack{+0.01 \\ -0.01}$&$0.34\substack{+0.01 \\ -0.01}$& $117 \substack{+1 \\ -1}$ &$1288\substack{+5 \\ -5}$\\
1282 &$17.84\substack{+0.08 \\ -0.08}$&$49.58\substack{+0.10 \\ -0.11}$&$-1.13\substack{+0.01 \\ -0.01}$&$54.34\substack{+0.03 \\ -0.03}$&$0.26\substack{+0.01 \\ -0.01}$&$1.05\substack{+0.01 \\ -0.01}$&$0.36\substack{+0.01 \\ -0.01}$& $119 \substack{+1 \\ -1}$ &$1340\substack{+26 \\ -28}$\\
1366 &$17.87\substack{+0.08 \\ -0.08}$&$49.68\substack{+0.10 \\ -0.10}$&$-1.14\substack{+0.01 \\ -0.01}$&$54.44\substack{+0.03 \\ -0.03}$&$0.24\substack{+0.01 \\ -0.01}$&$1.04\substack{+0.01 \\ -0.01}$&$0.35\substack{+0.01 \\ -0.01}$& $120 \substack{+1 \\ -1}$ &$1103\substack{+15 \\ -14}$\\
1421 &$17.91\substack{+0.10 \\ -0.09}$&$49.77\substack{+0.11 \\ -0.10}$&$-1.16\substack{+0.01 \\ -0.01}$&$54.53\substack{+0.05 \\ -0.04}$&$0.29\substack{+0.01 \\ -0.01}$&$1.05\substack{+0.01 \\ -0.01}$&$0.35\substack{+0.01 \\ -0.01}$& $122 \substack{+1 \\ -1}$ &$1029\substack{+18 \\ -16}$\\
1528 &$17.97\substack{+0.08 \\ -0.08}$&$49.92\substack{+0.10 \\ -0.11}$&$-1.17\substack{+0.01 \\ -0.01}$&$54.68\substack{+0.03 \\ -0.03}$&$0.19\substack{+0.01 \\ -0.01}$&$1.05\substack{+0.01 \\ -0.01}$&$0.35\substack{+0.01 \\ -0.01}$& $124 \substack{+1 \\ -1}$ &$890\substack{+26 \\ -28}$\\
1745 &$18.03\substack{+0.08 \\ -0.08}$&$50.07\substack{+0.09 \\ -0.09}$&$-1.18\substack{+0.01 \\ -0.01}$&$54.83\substack{+0.03 \\ -0.03}$&$0.15\substack{+0.01 \\ -0.01}$&$1.04\substack{+0.01 \\ -0.01}$&$0.34\substack{+0.01 \\ -0.01}$& $125 \substack{+1 \\ -1}$ &$614\substack{+5 \\ -5}$\\
1804 &$18.08\substack{+0.08 \\ -0.08}$&$50.20\substack{+0.09 \\ -0.09}$&$-1.20\substack{+0.01 \\ -0.01}$&$54.96\substack{+0.03 \\ -0.03}$&$0.14\substack{+0.01 \\ -0.01}$&$1.05\substack{+0.01 \\ -0.01}$&$0.34\substack{+0.01 \\ -0.01}$& $128 \substack{+1 \\ -1}$ &$626\substack{+4 \\ -4}$\\
1881 &$18.08\substack{+0.08 \\ -0.08}$&$50.20\substack{+0.09 \\ -0.09}$&$-1.19\substack{+0.01 \\ -0.01}$&$54.96\substack{+0.03 \\ -0.03}$&$0.12\substack{+0.01 \\ -0.01}$&$1.04\substack{+0.01 \\ -0.01}$&$0.33\substack{+0.01 \\ -0.01}$& $127 \substack{+1 \\ -1}$ &$535\substack{+3 \\ -2}$\\
1974 &$18.11\substack{+0.08 \\ -0.08}$&$50.26\substack{+0.09 \\ -0.09}$&$-1.20\substack{+0.01 \\ -0.01}$&$55.02\substack{+0.03 \\ -0.03}$&$0.11\substack{+0.01 \\ -0.01}$&$1.04\substack{+0.01 \\ -0.01}$&$0.33\substack{+0.01 \\ -0.01}$& $128 \substack{+1 \\ -1}$ &$479\substack{+2 \\ -2}$\\
2063 &$18.14\substack{+0.08 \\ -0.08}$&$50.34\substack{+0.09 \\ -0.09}$&$-1.20\substack{+0.01 \\ -0.01}$&$55.09\substack{+0.03 \\ -0.03}$&$0.10\substack{+0.01 \\ -0.01}$&$1.04\substack{+0.01 \\ -0.01}$&$0.33\substack{+0.01 \\ -0.01}$& $129 \substack{+1 \\ -1}$ &$444\substack{+2 \\ -2}$\\
2160 &$18.13\substack{+0.08 \\ -0.08}$&$50.33\substack{+0.09 \\ -0.09}$&$-1.20\substack{+0.01 \\ -0.01}$&$55.08\substack{+0.03 \\ -0.03}$&$0.08\substack{+0.01 \\ -0.01}$&$1.04\substack{+0.01 \\ -0.01}$&$0.32\substack{+0.01 \\ -0.01}$& $128 \substack{+1 \\ -1}$ &$375\substack{+1 \\ -1}$\\
 \enddata
 \tablecomments{Values in this table are calculated using an outflow launch time of $t_0=622$ days, $\nu_p$= 2.9 GHz, and $\epsilon_b = 0.001$.  For $\Gamma$ and $\beta$ we have accounted for the uncertainty in the launch date ($t_{0}=620\substack{+24 \\ -22}$ days).  Errors are calculating by propagating the RMS found in $\nu_p = 0.6$ GHz.}
\end{deluxetable*}

\section{Physical Properties of the Outflow}
\label{sec:params}

\begin{figure}
\begin{center}
    \includegraphics[width=.75\columnwidth]{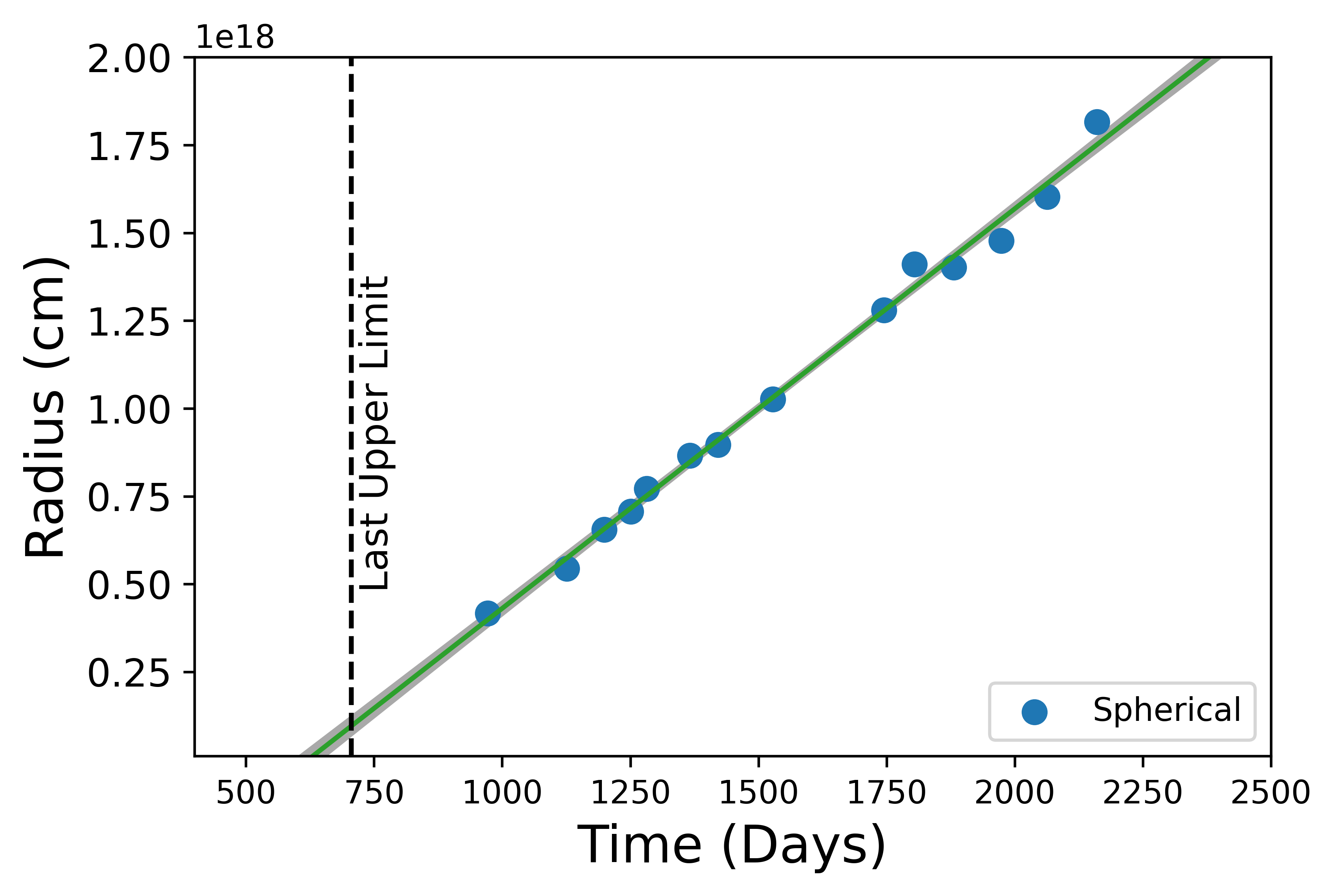}
    \end{center}
    \label{fig:disruption}
    \caption{Radius evolution of the outflow for the spherical geometry (Table~\ref{tab:equi}), assuming $\epsilon_{B}\approx 0.001$.  We fit a linear trend (i.e., free expansion) to these data to determine the launch time of the outflow (green line) and its uncertainty (grey shaded region marks the $1\sigma$ range). We find that $t_{\rm 0,sphere}=622\substack{+22 \\ -24}$ days relative to the time of optical discovery.  We also mark the time of the final non-detection observation at $705$ days for reference (vertical dashed line).}
\end{figure}

\subsection{Spherical Outflow}
\label{sec:spherical}

We begin by investigating the properties of the outflow in the conservative case of spherical geometry. We summarize the inferred physical parameters for all epochs in Table~\ref{tab:equi}. We find that the radius increases from ${\rm log}(R_{\rm eq})\approx 17.50$ to $\approx 18.13$ between 970 and 2160 d, corresponding to an average velocity of $\beta\approx 0.34$ over this time span. However, we find that the increase in radius is roughly linear, and fitting such an evolution with the launch time (i.e., time at which $R=0$) as a free parameter, we find $t_0\approx 620$ d (see Figure~\ref{fig:disruption}).  Thus, the physical evolution of the radius for a spherical outflow supports an outflow that was launched with a substantial delay of about 1.7 years relative to the discovery of optical emission.  The value of $t_0$ is about 85 days prior to the last radio upper limit at 705 days, but given the evolution of the light curve, the expected flux is well below the upper limit at that time.  The inferred outflow velocity using all epochs is $\beta\approx 0.33$. 

In the time span of 970 to 2160 days, the outflow kinetic energy increases by a factor of $\approx40$ from $E_K\approx 5.5\times 10^{48}$ to $2.1\times 10^{50}$ erg. The inferred ambient density declines from $n_{\rm ext}\approx 2.5$ to $\approx 1.2$ cm$^{-3}$, corresponding to a radial profile of $\rho(R)\propto R^{-1/2}$.  This is similar to the profile in M87*, Sgr A*, and Sw\,J1644+57, but shallower than the density profiles inferred around previous non-relativistic TDEs (see Figure \ref{fig:density} and citations therein).  

To conclude, in the conservative spherical outflow scenario we find that observed radio emission requires a delayed, mildly-relativistic outflow with a higher velocity and energy than in previous radio-emitting TDEs, and expanding into a relatively low density ambient medium.

\subsection{Off-Axis Relativistic Jet}
\label{sec:jet}

An alternative scenario for the delayed and steeply rising radio emission in \tde\ is an off-axis relativistic jet \citep{Matsumoto2023,Sfaradi2023}. The radio emission from an off-axis jet will be suppressed at early times by relativistic beaming, but will eventually rise rapidly  when the jet decelerates and spreads. Below we consider two off-axis jet model approaches, using the analytic formalism of \citet{Matsumoto2023}, which is an extension of the equipartition analysis in the spherical case from \S\ref{sec:equi}, and the numerical approach of \citet{Gill-Granot-18}, which can account for a more complex jet structure.  

\subsubsection{Generalized Equipartition Analysis}
\label{sec:matsumoto}

\citet{Matsumoto2023} generalized the equipartition analysis of \citet{Duran2013} to off-axis relativistic jets. The analysis showed that the Lorentz factor and observer viewing angle ($\theta_{\rm obs}$), cannot be determined independently and instead become degenerate along a trajectory of many minimal energy solutions. These solutions are further divided into on-axis and off-axis branches, and a relativistic source viewed off-axis can be disguised as an apparent non-relativistic one.

\begin{figure}[t!]
\begin{center}
\includegraphics[width=.6\columnwidth]{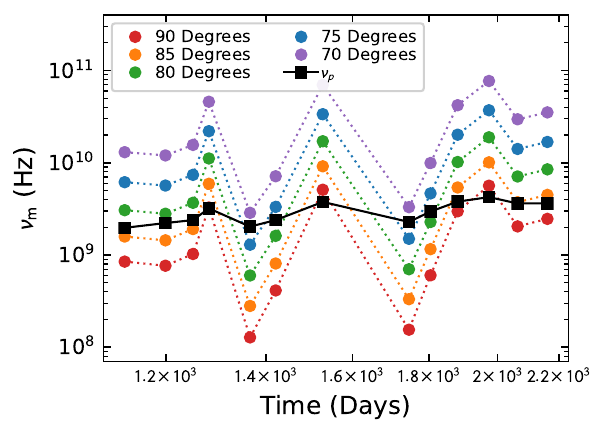}
\end{center}
\label{fig:matsumoto-num}
\caption{The characteristic synchrotron frequency $\nu_{\rm m}$ (Equation ~\ref{eq:num3}) for various angles for \tde. We assume a narrow jet with $\theta_{\rm j}=0.1$ and $\epsilon_e=0.1$ and $\epsilon_B=0.001$.  If we compare our derived values for $\nu_{\rm p}$ from Section \ref{sec:sed} (black points), we find our observations and inferred values for $\nu_m$ (Section \ref{sec:sed}) are consistent with a highly off-axis relativistic jet ($\theta_{\rm obs} \approx 80-90^{\circ}$).}
\end{figure}

\begin{figure}[t!]
\begin{center}
\includegraphics[width=.3\columnwidth]{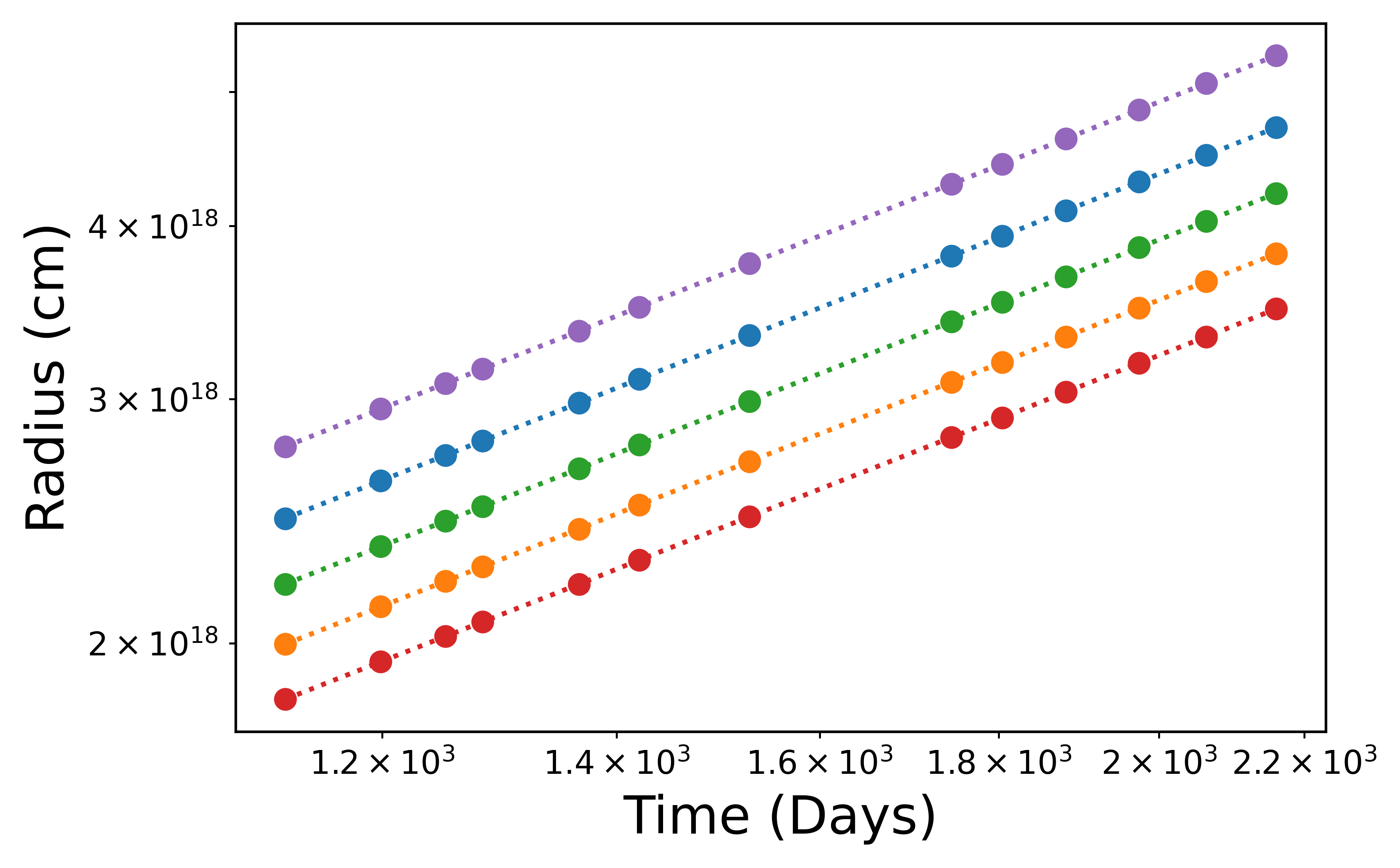}
\includegraphics[width=.3\columnwidth]{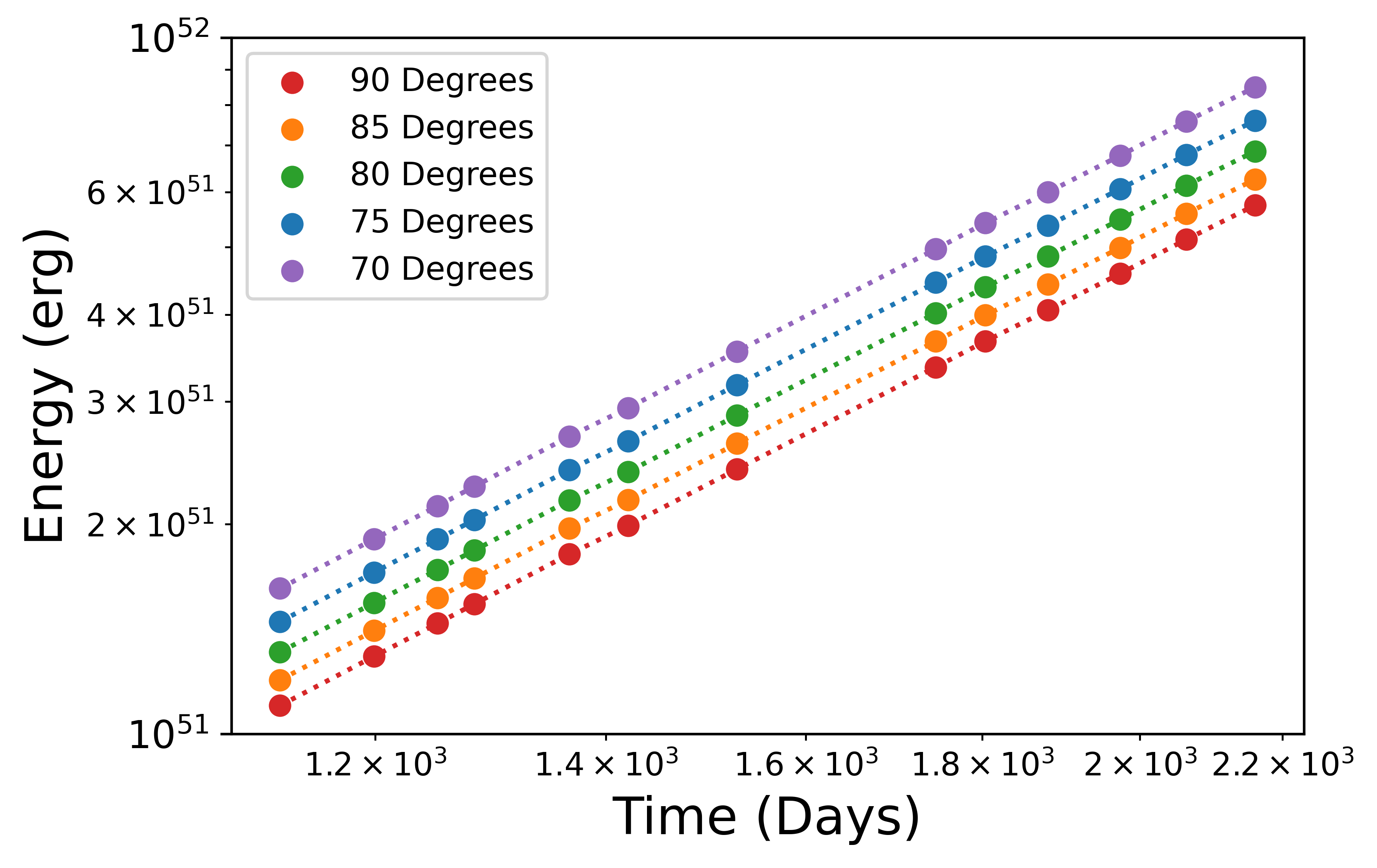}
\includegraphics[width=.3\columnwidth]{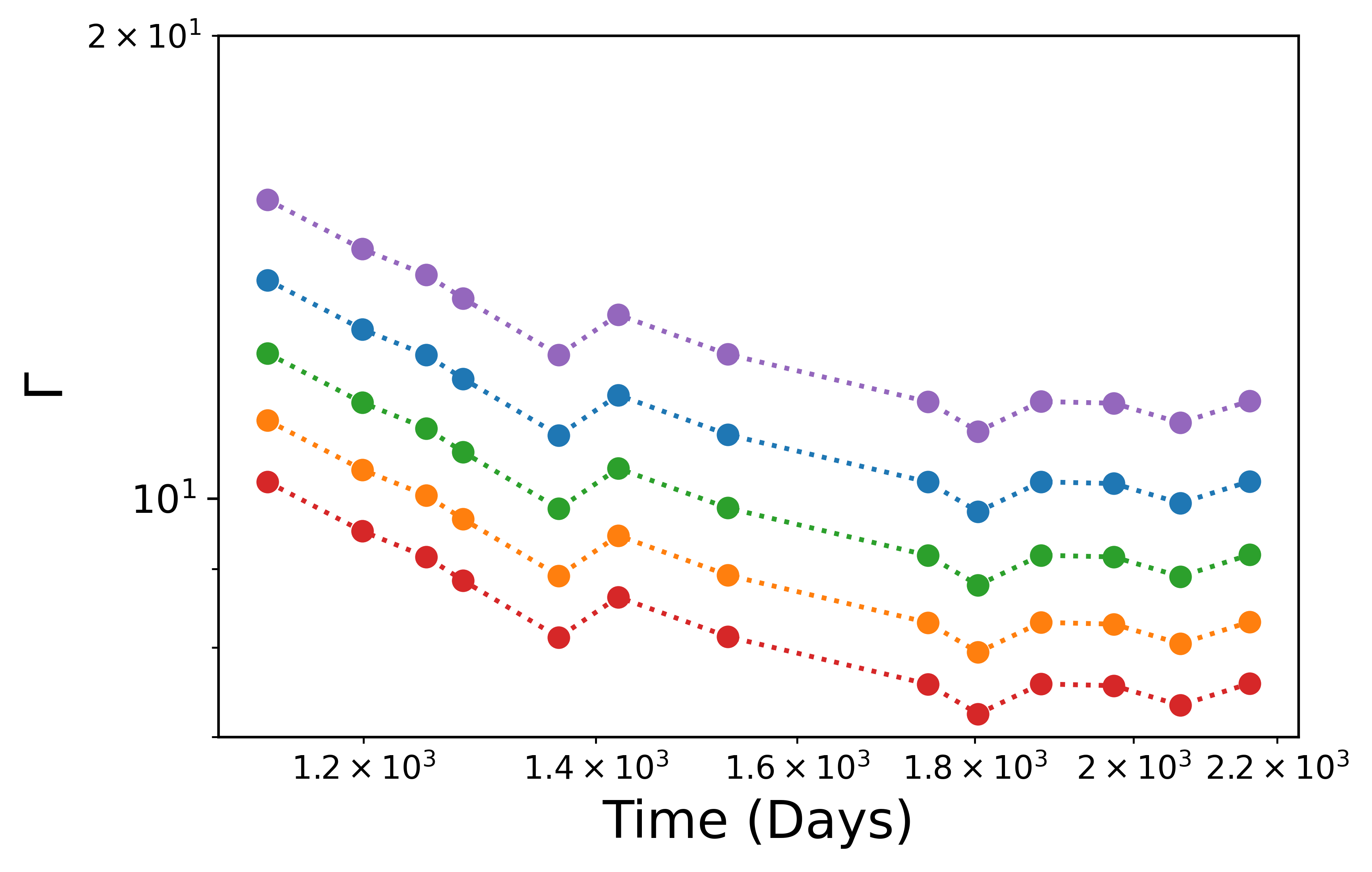}
\includegraphics[width=.3\columnwidth]{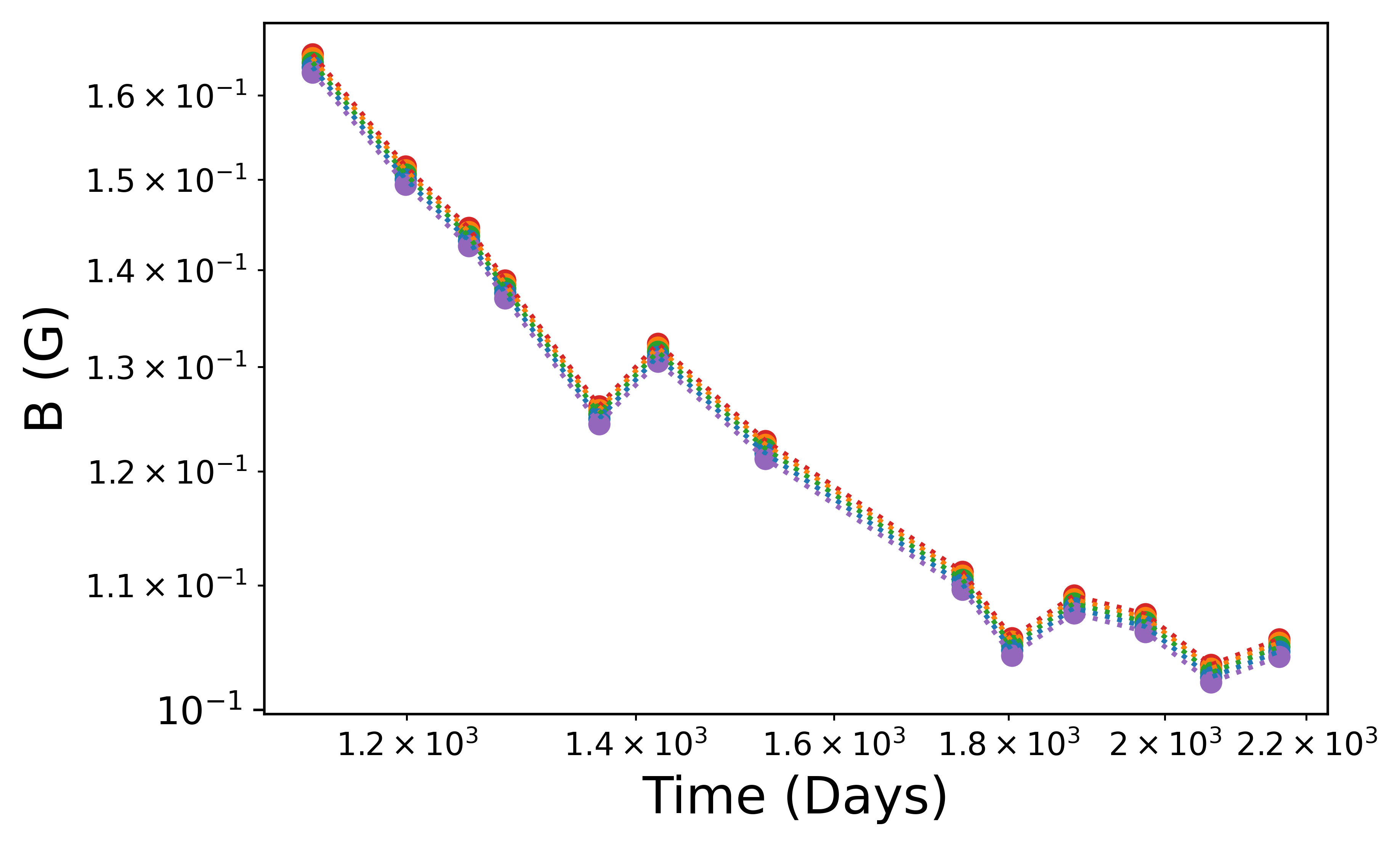}
\includegraphics[width=.3\columnwidth]{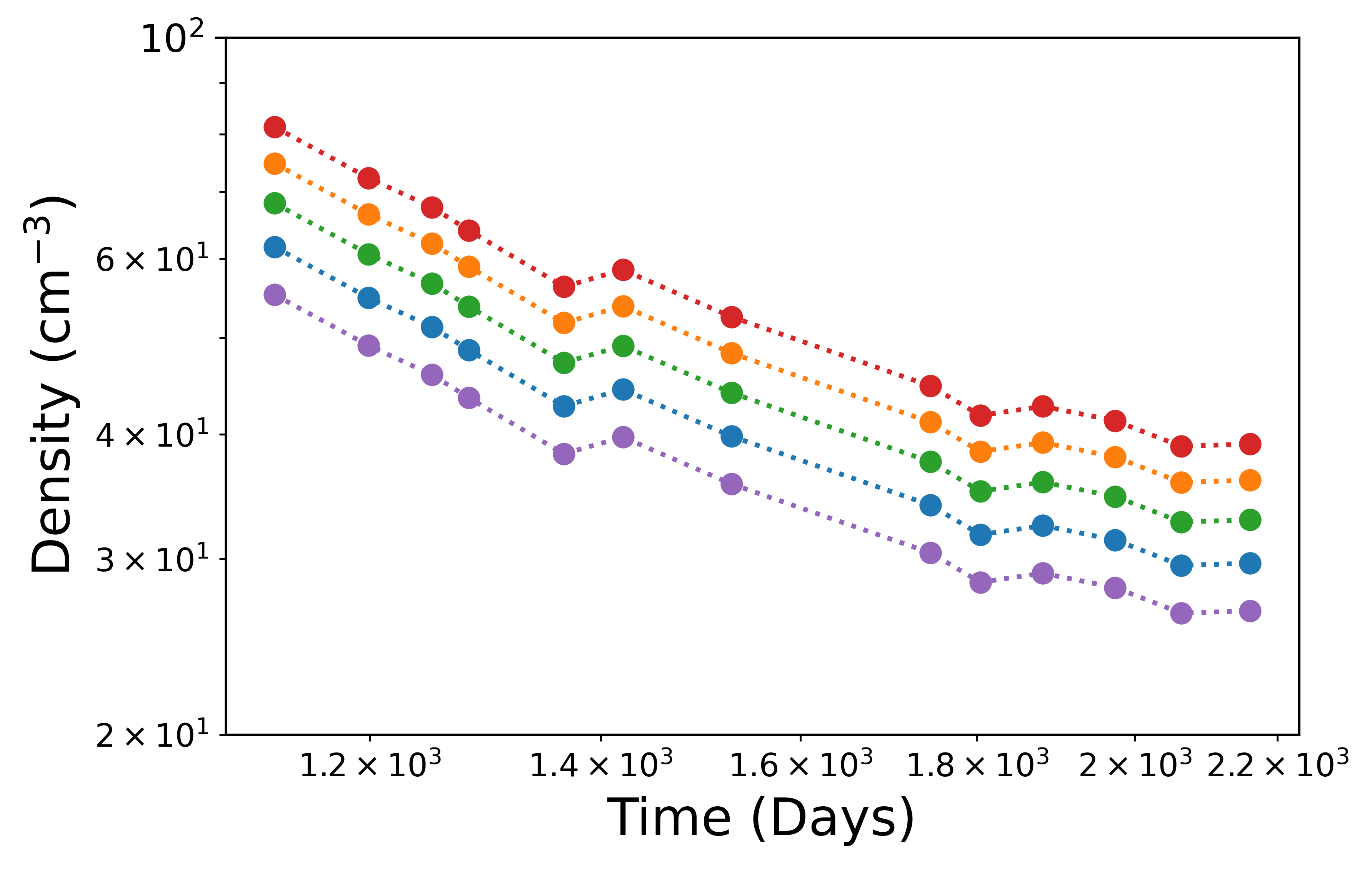}
\end{center}
\label{fig:matsumoto}
\caption{The radius, energy, $\Gamma$, magnetic field, and density calculated in \S\ref{sec:matsumoto} for an off-axis jet with several observer viewing angles ($\theta_{\rm obs}$).  Here, we are constrained by the limit on $\nu_m$ (Figure~\ref{fig:matsumoto-num}), and show off-axis angles of ($\theta_{\rm obs} \approx 80-90^{\circ}$).  We find for all cases that the radius and energy steadily increase over time, while the Lorentz factor, magnetic field, and density decrease over time.  The model is for $\theta_{\rm j}=0.1$, $\epsilon_B = 0.001$, and $\nu_p= 2.9$ GHz, and the energy includes a wide angle jet correction.}
\end{figure}

We note here, however, that a comparison between the spectral peak $\nu_{\rm p}$ and the characteristic frequency $\nu_m$ can actually provide an independent check on $\theta_{\rm obs}$. The latter is given by Equation 2 of \cite{Matsumoto2023}, which originally defines the observed peak frequency:
\begin{equation}
\nu_{\rm m} = \frac{\delta_D q_e B \gamma_m^2}{2\pi m_e c (1+z)}\ ,
\label{eq:num}
\end{equation}
where $q_e$ and $m_e$ are the electron charge and mass, respectively, $\delta_D$ is the relativistic Doppler factor, $B$ is the magnetic field at the source rest frame, and $\gamma_m$ is the Lorentz factor of electrons producing the synchrotron emission at $\nu_m$. Note that this equation simply means that the observed frequency is given by a Doppler boost. Since the Doppler factor and magnetic field are constrained by the equipartition analysis, Equation~\ref{eq:num} can be rewritten as:
\begin{align}
\nu_m & \approx 4.2\times10^{5}{\,\rm Hz\,}\left[\frac{\nu_{p,10}^{3/4}}{\eta^{5/12}}\left(\frac{t}{100{\,\rm d}}\right)^{-1/4}\right]\frac{f_A^{1/4}}{f_V^{1/4}}\gamma_m^{2}\ ,
\label{eq:num2}
\end{align}
where $\nu_{\rm p,10}=\nu_{\rm p}/10\,\rm GHz$, $\eta={\rm max}[1,\nu_{m}/\nu_{a}]$, $f_{\rm A}$ and $f_{\rm V}$ are the area- and volume-filling factors of the jet. Here we assume that the jet is relativistic ($\beta\approx 1$). The electrons'  minimum Lorentz factor is given by \citep[e.g.,][]{Sari1998}:
\begin{align}
\gamma_{m}&=\frac{m_p}{m_e}\left(\frac{p-2}{p-1}\right)\epsilon_{e}(\Gamma-1)\approx 46\,\bar{\epsilon}_{e,-1}(\Gamma-1)\ ,
\label{eq:gam_m}
\end{align}
where $\bar{\epsilon}_{e}\equiv 4\epsilon_{\rm e}\left(\frac{p-2}{p-1}\right)$ and $\bar{\epsilon}_{\rm e,-1}=\bar{\epsilon}_{\rm e}/0.1$. Substituting Equation~\ref{eq:gam_m} into Equation~\ref{eq:num2} we find:
\begin{align}
\nu_{m}&\approx 8.8\times10^{8}{\,\rm Hz\,}\left[\frac{\nu_{p,10}^{3/4}}{\eta^{5/12}}\left(\frac{t}{100{\,\rm d}}\right)^{-1/4}\right]\frac{f_{A}^{1/4}\bar{\epsilon}_{e,-1}^{2}(\Gamma-1)^{2}}{f_{V}^{1/4}} \ .
    \label{eq:num3}
\end{align}
Therefore, a direct measurement of $\nu_{\rm m}$ gives an independent constraint on $\Gamma$ and breaks the degeneracy between $\Gamma$ and $\theta_{\rm obs}$ in the equipartition analysis.

For \tde, we show in Figure \ref{fig:matsumoto-num} the inferred value of $\nu_m$ from Equation~\ref{eq:num3} for a range of $\theta_{\rm obs}=70-90\,\rm deg$. We assume a narrow jet geometry with $f_{A}=f_{V}=(\Gamma\theta_{j})^2$ and $\theta_{j}=0.1\,(\approx5.7^{\circ})$ (see Appendix B in \citealt{Matsumoto2023}), and take into account the deviation from equipartition with $\epsilon_{e}=0.1$ and $\epsilon_{B}=10^{-3}$. In addition, we fix $\eta=1$, which is justified given our finding that $\nu_{a}\approx \nu_{m}$. We find that only viewing angles of $\theta_{\rm obs} \gtrsim 80^{\circ}$ have predicted values of $\nu_m$ in agreement with the observed value; smaller off-axis angles of even $\approx 70^\circ$ predict $nu_m\gtrsim 10$ GHz, in conflict with the observed SEDs.  We therefore conclude that if AT2018hyz harbors an off-axis jet, then its viewing angle is $\approx 80-90^{\circ}$.  

Using this information, we derive the physical parameters of the jet, including $R$, $E_K$, $\Gamma$, $B$, and $n_{\rm ext}$; see Figure~\ref{fig:matsumoto}.  We find that $R$ increases from $\approx 2\times 10^{18}$ to $\approx 4\times 10^{18}$ cm, $E_K$ increases from $\approx 10^{51}$ to $\approx 6\times 10^{51}$ erg, while $\Gamma$ decreases from about 10 to 8, and the density decreases from $\approx 80$ to $\approx 40$ cm$^{-3}$.



\subsubsection{Off-Axis Jet with Steep Angular Structure -- Forward modeling}
\label{sec:psjet}

We next test the off-axis jet scenario with forward modeling of the emission, taking into account the jet hydrodynamics and the effect on the temporal and spectral evolution self-consistently. Furthermore, the unlike the equipartition analysis which assumes a ``top-hat'' (uniform) jet, here we consider the possibility of a jet with steep gradients in its energy per unit solid angle, 
$dE/d\Omega\equiv\epsilon(\theta)=\epsilon_c\Theta^{-a}$, and bulk Lorentz factor, $\Gamma(\theta) = 1+(\Gamma_c-1)\Theta^{-b}$, where
$\Theta(\theta) = \sqrt{1 + (\theta/\theta_c)^2}$, $\theta$ is the polar angle, $\theta_c$ is the jet core angle and $\epsilon_c$ and $\Gamma_c$ are the core values of the profiles defined at $\theta=0$
\citep[e.g.][]{Gill-Granot-18,Beniamini-Granot-Gill-20,Beniamini-Gill-Granot-22}. The jet propagates in an external medium with density $\rho_{\rm ext}(R)=n_{\rm ext}(R)m_p = AR^{-k}$, and the resulting afterglow synchrotron emission is calculated assuming a thin-shell geometry that ignores any radial profile of the shocked swept-up medium behind the external forward shock. Furthermore, for computational convenience, we adopt locally spherical dynamics that evolves each part of the outflow assuming spherical dynamics given the local isotropic-equivalent kinetic energy $E_{k,\rm iso}(\theta)=4\pi\epsilon(\theta)$ and bulk $\Gamma(\theta)$. Lateral spreading of the flow, which becomes important as the initially relativistic jet is sufficiently decelerated \citep{Rhoads-99,Sari+99,Granot-Piran-12}, is not included in the model \citep[but see, e.g.,][for a discussion on lateral expansion]{Kumar-Granot-03,vanEerten-MacFadyen-12,Duffell-Laskar-18}. 
The viewing geometry is parameterized using $q\equiv\theta_{\rm obs}/\theta_c$, where $\theta_{\rm obs}$ is the observer viewing angle. The final emission is obtained by integrating over the equal-arrival-time-surface that accounts for the common time of arrival of radiation emitted at different angles away from the line-of-sight and at different radii/lab-frame times.

\begin{figure}[t!]
    \centering
    \includegraphics[width=0.45\linewidth]{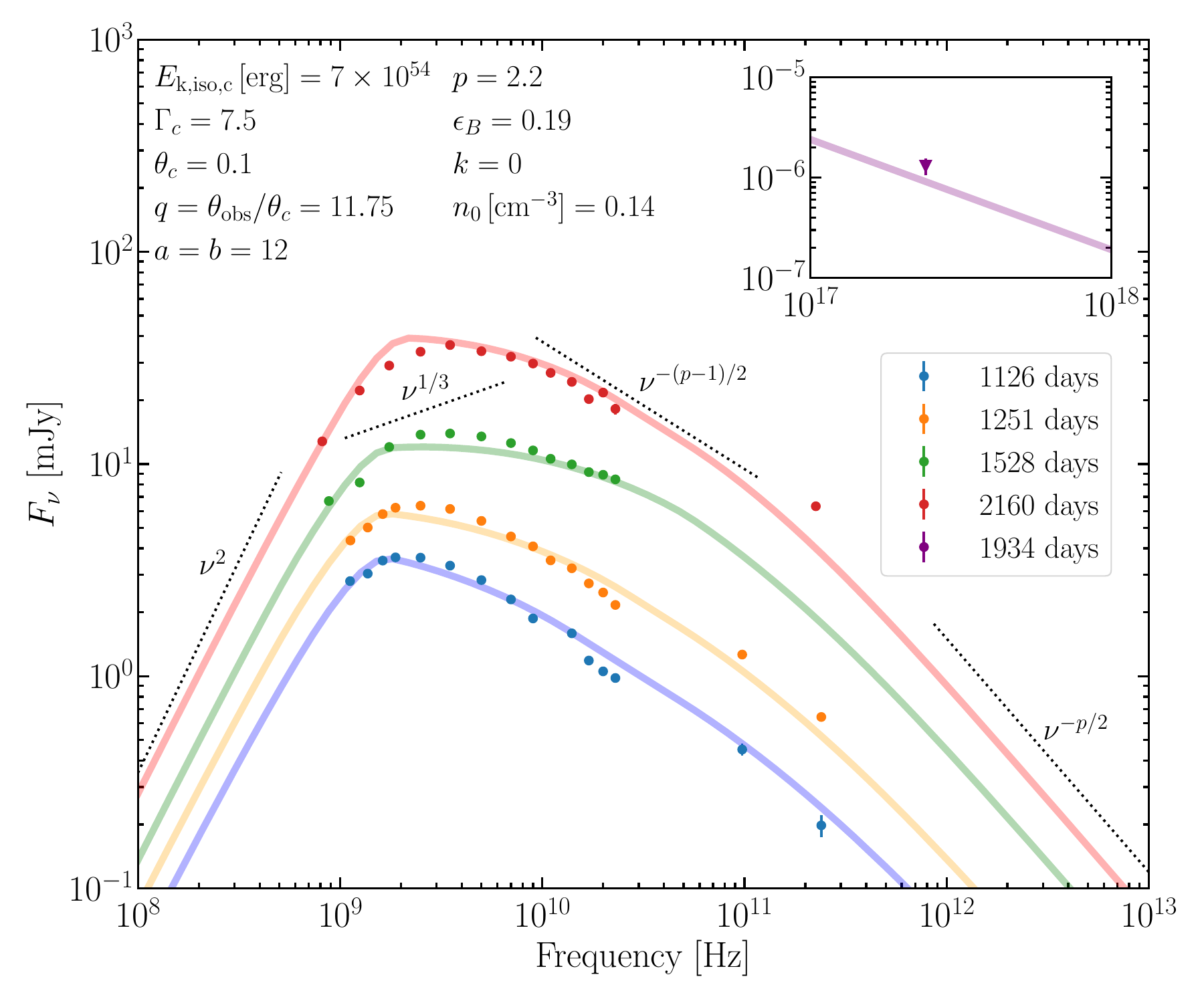}
    \includegraphics[width=0.45\linewidth]{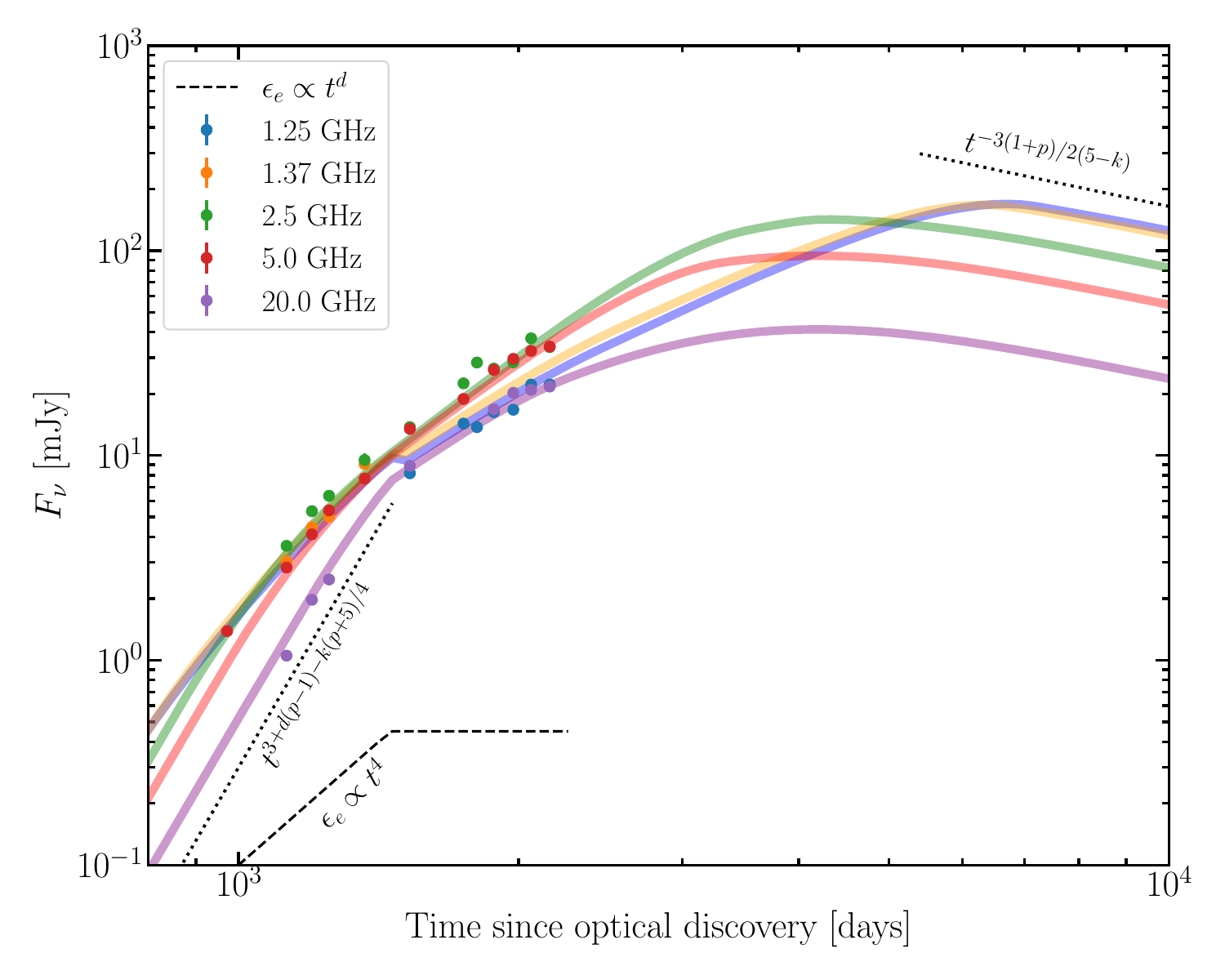} \\
    \caption{Off-axis structured jet afterglow model comparison. 
    \textit{Left:} The spectrum for \tde at different epochs, with the standard spectral slopes shown with dotted line segments. These are not generally obtained here due to overlapping spectra from neighboring angles that have a different $E_{\rm k,iso}(\theta)$ and bulk $\Gamma(\theta)$ as prescribed by the steep angular structure of the flow. All of the parameters of the model are presented in this panel. The inset shows the X-ray upper limit; note that axes units are the same as for the larger figure. Increasing the electron cooling $\gamma_c$ by a factor of $\sqrt{2}$ (i.e. $\gamma_c\to\sqrt{2}\gamma_c$) in our model yields a better fit to the sub-mm data. \textit{Right:} The light curves at different observer-frame frequencies, with the contribution from the main and counter jets included. The black dotted line segments show the expected temporal slopes at early and late times. The assumed temporal evolution of $\epsilon_e$ is shown with a dashed line, where it changes by a factor $\sim 2$ over the duration of the observations. 
 }
    \label{fig:structure-jet-model-fits}
\end{figure}

In Figure~\ref{fig:structure-jet-model-fits} we show the resulting afterglow model for a constant density external medium ($k=0$), compared to both sample SEDs and light curves.  Our model includes an {\it ad-hoc} temporal evolution of $\epsilon_e\propto t^4$ in the first few hundred days of evolution (dashed line in the light curve panel of Figure~\ref{fig:structure-jet-model-fits}), where the value of $\epsilon_e$ changes by a factor of $\approx 2$. This is required to explain the mild hardening of the SED above the spectral peak at later epochs. A fixed $\epsilon_e$ yields much softer high-energy model spectra above the spectral peak (at $\nu\approx2$\,GHz) that disagree with observations.

The key finding is that the model requires a large off-axis viewing angle, with $q\approx 12$, in good agreement with the simple top-hat calculation in \S\ref{sec:matsumoto}. This, coupled with the steep gradient in the angular structure, with $a\approx b\approx 12$, results in the emission being completely dominated by material within the jet core at all times.  We discuss the dynamical evolution of the jet angle ($\theta_F$) and associated parameters in Appendix \ref{sec:appjet}.

\section{Discussion and Comparison to Other Radio-Emitting TDEs}
\label{sec:disc}

It has recently been shown that $\sim40\%$ of all optically detected TDEs with no prior radio emission detected at early times will have detectable radio emission hundreds of days post-disruption \citep{Cendes2024}.  At its discovery, and during the period covered in \citet{p2}, \tde\ had a comparable luminosity to many of this population of late-emitting TDEs ($10^{39}$ erg/s at $t_d \approx 1000$ days).  However, \tde\ is exceptional if we consider its continued evolution since, with its continued rise in luminosity to that of the jetted TDE population at these time scales ($\gtrsim 10^{40}$ erg/s at $t_d \approx 2200$ days).  Such a rapid rise, combined with a rise at late time scales, is unprecedented in previous TDEs.

ASASSN-15oi exhibits two episodes of rapid brightening, at $\approx 200$ and $\approx 1400$ d \citep{Horesh2021}.  While the second brightening is not well characterized temporally or spectrally, it has a comparable rise rate and luminosity to \tde.  We speculate that it may be due to a delayed outflow with similar properties to that of \tde, including a delay of several hundred days, which would make it distinct from the first peak in the ASASSN-15oi light curve.

\begin{figure}
\begin{center}
    \includegraphics[width=.9\columnwidth]{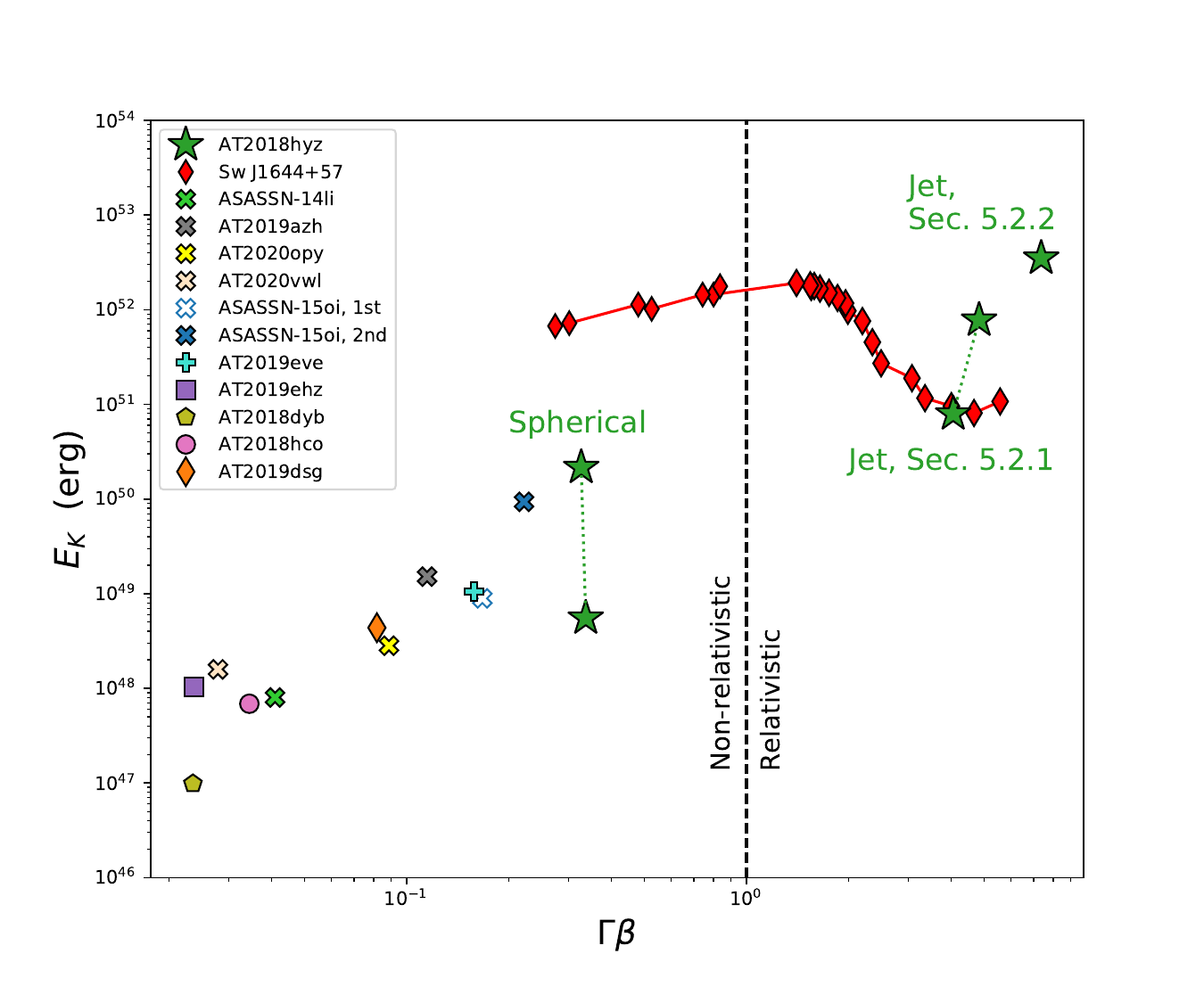}
    \end{center}
    \label{fig:Ev}
    \caption{The energy/velocity for \tde~ spherical case and jetted cases outlined in Section \ref{sec:jet}. For simplicity, we include only data from our first and last observations for the spherical case and jet in Section \ref{sec:matsumoto}; note in these cases energy increases over time.  We include for reference a selection of non-relativistic TDEs assuming a spherical outflow \citep{Cendes2021b,Stein2020,Alexander2016, Alexander2017, Anderson2019,Goodwin2022}, showing the highest energy found for each TDE \citep[as described in ][]{p2}.  We also show data from Sw\,J1644+57 over time, which launched a relativistic jet \citep{Zauderer2011,Cendes2021}.  We find that in the spherical case (Section \ref{sec:spherical}, \tde~ is more energetic than the non-relativistic outflow TDEs, and has a higher velocity.  In the case of an off-axis relativistic jet, we find the velocity and energy similar to that seen in Sw\,J1644+57 or higher.}
\end{figure} 

\subsection{Outflow Kinetic Energy and Velocity}

In Figure~\ref{fig:Ev} we plot the kinetic energy and velocity ($\Gamma \beta$) of the delayed outflow in \tde\ in comparison to previous TDEs for which a similar analysis has been carried out, using the highest energy inferred in those sources \citep{Cendes2021b,Stein2020,Alexander2016, Alexander2017, Anderson2019,Goodwin2022,Zauderer2011,Cendes2021,Hajela2024,Cendes2024}.  We include both the spherical (Section \ref{sec:spherical}) and $90^{\circ}$ off-axis jet models (Section \ref{sec:matsumoto} and Section \ref{sec:psjet}), plotting the first and last data point of our model in each case.  

We find that in our last observation for the spherical case the energy is a factor of 2 times larger and the velocity is $\approx 3$ times faster than in previous non-relativistic TDEs, with the exception of the second outflow of ASASSN-15oi.  Thus, if \tde\ was created by a delayed outflow, it would have significantly higher velocity and energy than what has been seen in all other TDEs with a delayed outflow to date \citep[][ and references therein]{Cendes2024}. 
 The exception to this is the second flare of ASASSN-15oi, which rose between 576 and 1741 days post-disruption by a factor of $\sim3000$, or a corresponding power law ($F_\nu \propto t^\alpha$) with $\alpha\approx 4.5$ \citep{Horesh2021,Hajela2024}.  If we assume a delayed launch for the second flare of ASASSN-15oi of 576 days post disruption (the date of last observation before the second flare was observed), we find $\beta \approx 0.17$, or $\sim1.5\times$ less than the outflow as modeled in the spherical case of \tde.  We also find the energy in the second flare of ASASSN-15oi is comparable to that found in \tde\ \citep{Hajela2024}.  We note that due to a lack of data sampling in the ASASSN-15oi flare, the exact start and end date of the flare is less known, and its velocity could be higher \citep[though this is common in the light curves of several late-time TDE flares; see][for additional examples]{Cendes2024}.
 
 If the outflow in \tde{} is collimated, then its velocity and energy is similar to that inferred from powerful jetted TDEs such as Sw\,J1644+57.

\subsection{Circumnuclear Density}

\begin{figure}
\begin{center}
    \includegraphics[width=.9\columnwidth]{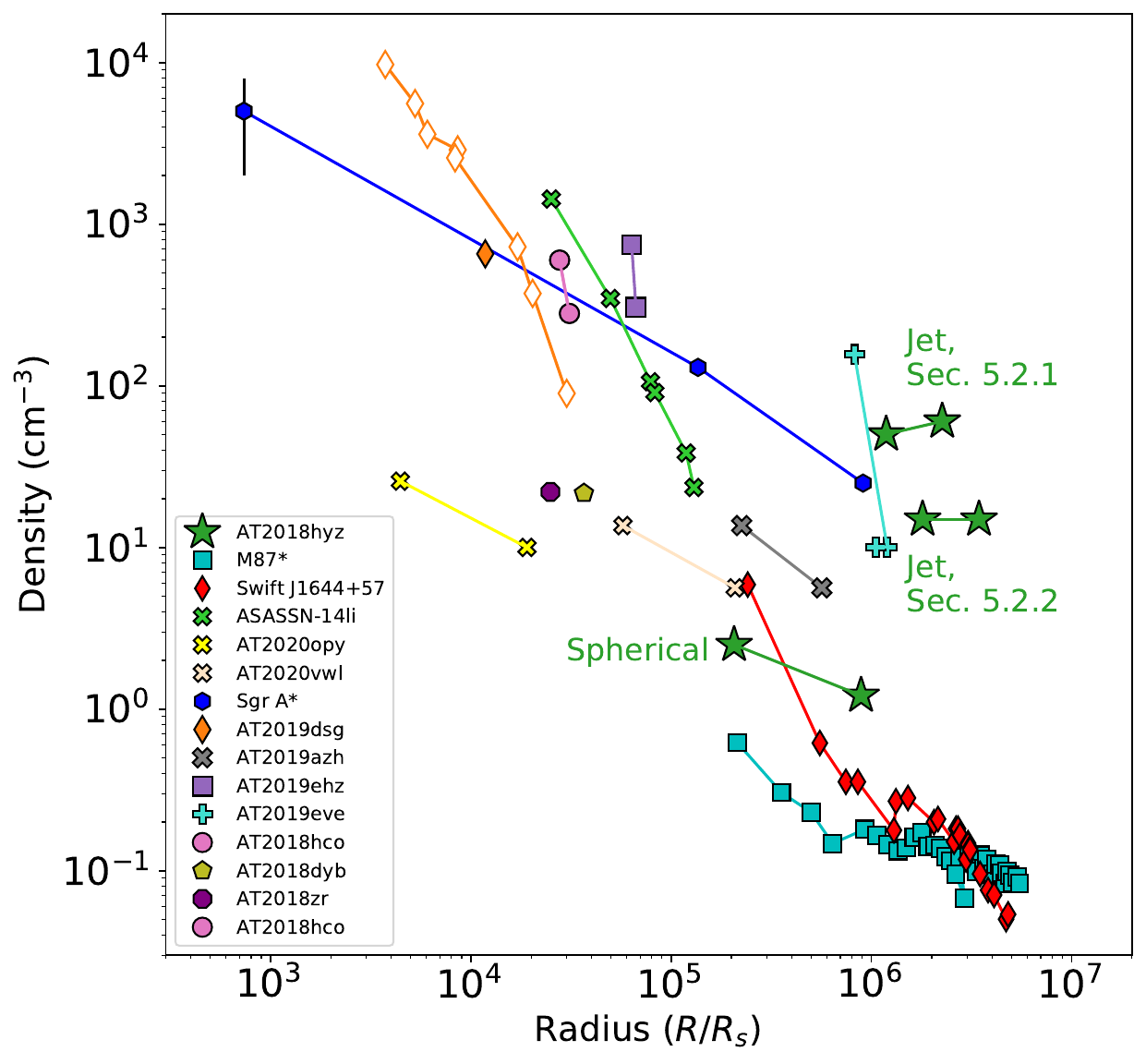}
    \end{center}
    \label{fig:density}
    \caption{The circumnuclear density profile derived from various TDEs including \tde, normalized to the Schwarzschild radius of the SMBH at each host galaxy's center, for both the spherical and off-axis jet models.  We include data from only the first and last observations for reference.  In the spherical case, \tde's host galaxy is lower density, similar to that seen in the jetted TDE SwJ1644+57 \citep{Berger2012,p2,Eftekhari2018} and M87 \citep{Russell2015}.  In the jetted case, we find a density environment similar to that seen in the Milky Way with both inferred values of $\epsilon_{B}$ in our modeling \citep{Baganoff2003,Gillessen2019}, and a fairly constant density environment, unusual for TDEs.  We also include non-relativistic TDEs (e.g.~ASASSN-14li, \citep{Alexander2016}, AT2019dsg \citep{Cendes2021b}, AT2019azh \citep{Goodwin2022}, ASASSN-15oi \citep{Horesh2021}).  Other than for AT2019dsg and Sw\,J1644+57, where the cooling break is detected, we assume equipartition in all radio TDEs, so their densities are lower limits.}
\end{figure}

In Figure~\ref{fig:density} we plot the ambient density as a function of radius (scaled by the Schwarzschild radius) for \tde\ and previous radio-emitting TDEs.  Here we use $M_{\rm BH}\approx 5.2\times 10^{6} M_{\odot}$ for \tde, as inferred by \citet{Gomez2020}.  We find that in the spherical case the density decreases with radius, and is consistent with the densities and circumnuclear density profiles of previous TDEs, including Sw\,J1644+57.  Crucially, we do not infer an unusually high density, which might be expected if the radio emission was delayed due to rapid shift from low to high density.

In the case of a collimated, off-axis jet, we find in both cases a density that is roughly constant and roughly a factor of 2 higher than in the spherical case.  This density is similar to what has been measured in other TDEs \citep{p2,Cendes2021b,Alexander2016}, and in our own Milky Way \citep{Baganoff2003,Gillessen2019}, but is a much more shallow density profile than seen before in a TDE over a $\sim1200$ day period.



\section{Conclusions}
\label{sec:conc}

We presented the discovery of continued late- and rapidly-rising radio/mm emission from \tde\ starting at about 970 d post optical discovery, and extending to at least 2160 days. The radio emission is rising at all frequencies and is more luminous than all previous non-relativistic TDEs, and approaching that seen in the relativistic TDE Sw\,J1644+57.  The multi-frequency data indicates that the peak frequency of emission is relatively stable, with an increase of peak flux density dominating the evolution of this source, increasing by an order of magnitude.  

Our analysis suggests two possibilities for the outflow:
\begin{itemize}
    \item A spherical outflow launched $\approx 620$ d after optical discovery with a velocity of $\beta\approx 0.3$, and $E_K\gtrsim 10^{50}$ erg in our most recent observations.  This is in excess of all previous non-relativistic TDEs.
    \item A highly off-axis ($\approx 80-90^\circ$) relativistic jet, with either a ``top-hat" (uniform) jet and a jet with steep angular structure.  In both cases we find the jet would have a Lorentz factor of $\Gamma\sim 8$, $E_K\approx 10^{52}$ erg, and a relatively flat density profile.  Our models indicate if \tde\ is an off-axis jet, we can expect a turnover in the light curve at higher frequencies ($\lesssim5$ GHz) at $\approx$3,000 days post-disruption (early 2027), and much later ($t_d \approx$6,000 days) at the peak frequency of emission.
\end{itemize}

We find that \tde is a unique TDE even within the population of TDEs with delayed radio emission, and future observations should allow us to distinguish between these scenarios.  We have planned continued multi-frequency observations of \tde\ to monitor the on-going evolution of the outflow and of the circumnuclear medium, including VLBI.

\begin{acknowledgments}
We thank Emil Polisensky for his assistance with VLITE data, and the VLA, MeerKAT, and SMA observatory staff for all their assistance.  This research was supported in part by grant NSF PHY-2309135 to the Kavli Institute for Theoretical Physics (KITP).  The Berger Time-Domain Group at Harvard is supported by NSF and NASA grants. This paper makes use of the following ALMA data: ADS/JAO.ALMA\#2021.1.01210.T.  ALMA is a partnership of ESO (representing its member states), NSF (USA) and NINS (Japan), together with NRC (Canada), MOST and ASIAA (Taiwan), and KASI (Republic of Korea), in cooperation with the Republic of Chile. The Joint ALMA Observatory is operated by ESO, AUI/NRAO and NAOJ. The National Radio Astronomy Observatory is a facility of the National Science Foundation operated under cooperative agreement by Associated Universities, Inc.  N. Velez was supported by a SOS award from NRAO. The scientific results reported in this article are based in part on observations made by the Chandra X-ray Observatory, which is operated by the Smithsonian Astrophysical Observatory for and on behalf of the National Aeronautics Space Administration under contract NAS8-03060.  The MeerKAT telescope is operated by the South African Radio Astronomy Observatory, which is a facility of the National Research Foundation, an agency of the Department of Science and Innovation.  The Australia Telescope Compact Array is part of the Australia Telescope National Facility (https://ror.org/05qajvd42) which is funded by the Australian Government for operation as a National Facility managed by CSIRO.  The Submillimeter Array is a joint project between the Smithsonian Astrophysical Observatory and the Academia Sinica Institute of Astronomy and Astrophysics and is funded by the Smithsonian Institution
and the Academia Sinica. We recognize that Maunakea, where the SMA is sited, is a culturally important site for the indigenous Hawaiian people; we are privileged to study the cosmos from its summit.
\end{acknowledgments}

\appendix
\label{sec:appjet}

\begin{figure}[t!]
    \centering
    \includegraphics[width=0.45\linewidth]{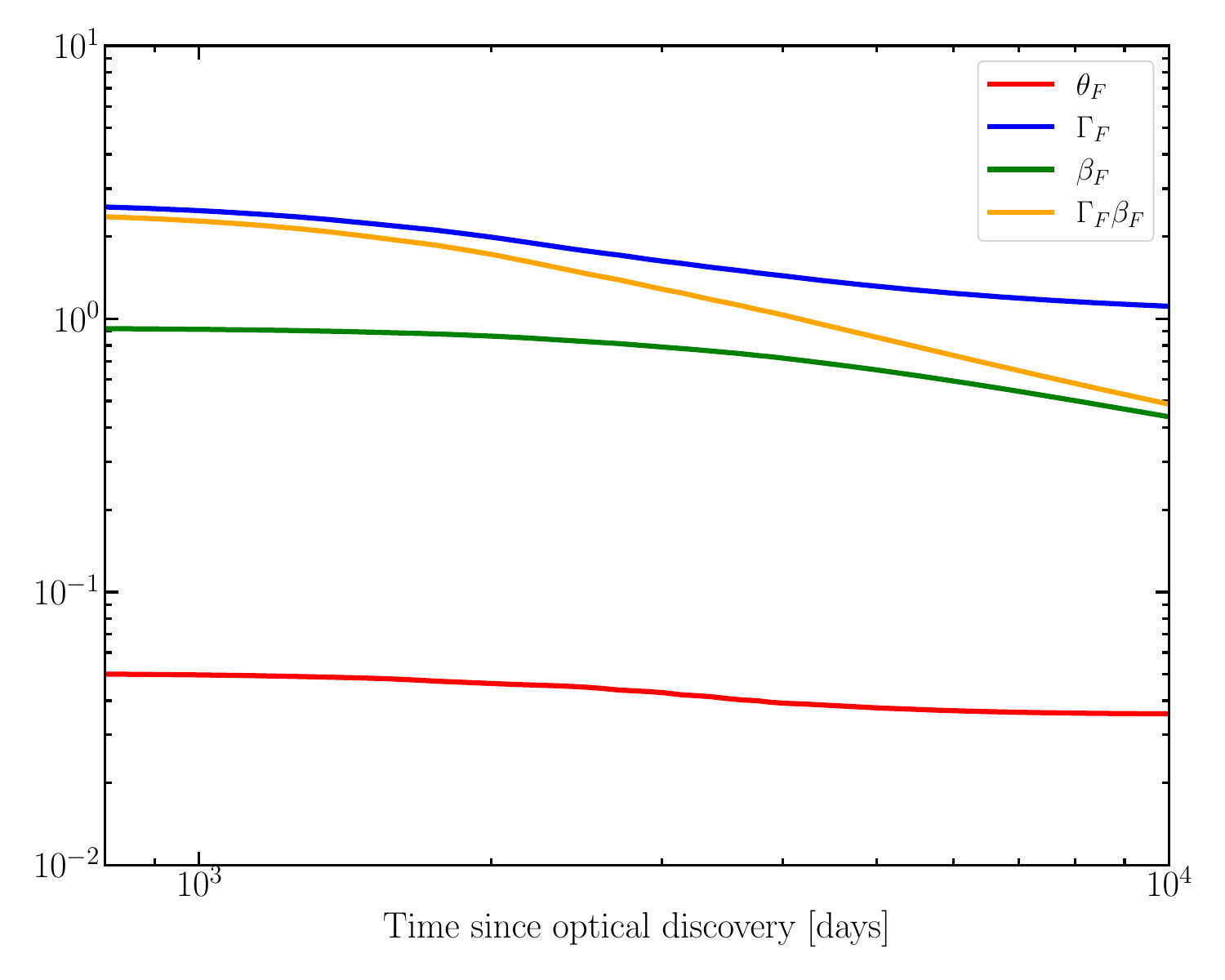}
    \caption{Dynamical evolution of the angle ($\theta_F$) at which emission from material dominates the observed 
    flux at 100\,GHz. The different quantities show its bulk $\Gamma_F = \Gamma(\theta_F)$, non-dimensional radial velocity $\beta_F$, 
    and four speed $\Gamma_F\beta_F$.
 }
    \label{fig:structure-jet-model-fitsa}
\end{figure}

In \S\ref{sec:psjet}, different observed frequencies are dominated at a given observed time by different angles within the jet core.  This is unlike the case for less steep jets \citep{Beniamini-Granot-Gill-20}.  Figure~\ref{fig:structure-jet-model-fitsa} shows the angle $\theta_F$ which is defined as the angle that dominates the observed flux at 100\,GHz at a given time. For a different observed frequency, this angle and its temporal evolution will be different. The dynamical evolution of the material at this angle is also shown using its bulk $\Gamma_F = (1-\beta_F^2)^{-1/2}$ and propagation four speed, $\Gamma_F\beta_F$. At early times, the flux is dominated by material at $\theta_{F,0}=0.5\theta_c$, which decelerates during the observations at \citep{Beniamini-Granot-Gill-20}:
\begin{eqnarray}
    R_{\rm dec}(\theta_{F,0}) &=& \left[\frac{(3-k)E_{\rm k,iso}(\theta_{F,0})}{4\pi Ac^2\Gamma_0^2(\theta_{F,0})}\right]^{1/(3-k)}
    \approx 6.6\times10^{18}\,{\rm cm} \\
    t_{\rm dec}(\theta_{F,0}) &=& \frac{R_{\rm dec}(\theta_{F,0})(1+z)}{c\beta_{F,0}}[1-\beta_{F,0}\cos(\theta_c\{q-\theta_{F,0}/\theta_c\})] 
    \approx  1.6\times10^3\,{\rm days}\,,
\end{eqnarray}
where $\beta_{F,0}=\beta(\theta_{F,0}) = [1-\Gamma(\theta_{F,0})^{-2}]^{1/2}$. After this point, angles smaller than $\theta_{F,0}$ start 
to dominate the flux.

Over the observed epoch, the afterglow emission is produced when the dominant angle is in the trans-Newtonian regime, such that emission from neighboring angles is not strongly beamed away and therefore contributes appreciably to the received flux. In addition, due to the steep gradient in the energy profile within the core, the overlapping 
spectral breaks from neighboring angles lead to significantly broadened spectral breaks in the net emission. At these times, the angle dominating the emission is such that $\nu_m(\theta_{\rm F})\sim\nu_a(\theta_{\rm F})$, and the spectral peak is then at $\nu_a\sim \nu_m$ (left panel of Fig.\,\ref{fig:structure-jet-model-fits}). The spectral shape below and above the peak is strongly affected by the steep angular structure. Below the peak, the non-standard spectral shape (which is neither $\nu^2$ nor $\nu^{1/3}$) is produced by the overlap of spectral peaks occurring at $\nu_a > \nu_m$ from the different angles with $\theta>\theta_F$. In contrast, the non-standard spectral shape (which is neither $\nu^{1/3}$ nor $\nu^{-(p-1)/2}$) above the break is given by the overlapping spectral peaks at $\nu_m$ from angles $\theta<\theta_F$. As a result, $\nu_a$ is structure broadened at $\sim 1$ GHz, $\nu_m$ forms a broad peak at $\sim 10$ GHz, and $\nu_c$ is broadened at $\sim 100$ GHz. The implication is that the spectrum is modified as compared to a one-zone synchrotron emission model. 

The light curve at different frequencies is shown in the right panel of Fig.\,\ref{fig:structure-jet-model-fits}. The temporal scaling during the initial rising phase pre-deceleration, when $\Gamma=\Gamma_0(\theta_F)$ and $\epsilon_e\propto t^d$, for $\nu_m<\nu<\nu_c$ is given by $d\ln F_\nu/d\ln t = 3+d(p-1)-k(p+5)/4$. Post-deceleration, the temporal scaling of the light curve below and above the spectral peak becomes non-standard and more complex due to overlapping of emission from different angles. 
At later times, different frequencies show peaks at different times. This is indicative of a spectral change rather than a geometric one. Here the spectral break at each angle, $\nu_m \propto t^{(4k-15)/(5-k)}$ (with post-deceleration scaling for a Newtonian flow), crosses the observed frequency and that causes the rising light curve, when $\nu_{\rm obs}<\nu_m<\nu_c$, to transition into a declining one, when $\nu_m<\nu_{\rm obs}<\nu_c$. 

With the given scaling of $\nu_m$ with time, it also means that higher frequencies will show a light curve peak earlier than lower ones. Due to the contribution of different angles at any given time, the actual temporal evolution of the critical frequencies is affected by the angular structure and is different from that obtained locally. After the light curve peak, all frequencies show the same rate of decline in flux density, 
which means that all are located on the same power-law segment of the synchrotron spectrum, which in this case is $\nu_m<\nu<\nu_c$. At 
such late times, the flow enters the deep-Newtonian regime which sees a reduction in the total number of synchrotron-producing 
shocked electrons that can be accelerated to a power-law energy distribution above a certain minimal particle Lorentz factor 
$\gamma_m\approx\sqrt{2}$ \citep[e.g.][]{Beniamini-Gill-Granot-22}. The temporal scaling in this regime is given 
by $d\ln F_\nu/d\ln t = -3(p+1)/2(5-k)$.

\appendix
\label{sec:app}

\startlongtable
\begin{deluxetable*}{lcclll}
\label{tab:obs}
\tablecolumns{6}
\tablecaption{Radio and Millimeter Observations of \tde}
\tablehead{
\colhead{Date} & \colhead{Observatory} & \colhead{Project} &
\colhead{$\delta t$\,$^a$} & \colhead{$\nu$} & \colhead{$F_\nu$\,$^b$} \\
\colhead{} & \colhead{} & \colhead{} & \colhead{(d)} & \colhead{(GHz)} & \colhead{(mJy)}
}
\startdata
2021 Nov 12$^c$ &VLA & VLITE & 1126 & 0.34 & <4.4\\
& & 21B-357 &  & 1.37 & 3.043$\pm$0.054\\
& &  &  & 1.62 & 3.505$\pm$0.084\\
& &  &  & 1.88 & 3.629$\pm$0.065\\
& &  &  & 2.5 & 3.618$\pm$0.047\\
& &  &  & 3.5 & 3.317$\pm$0.040\\
& &  &  & 5 &2.836$\pm$0.029\\
& &  &  &  7&2.299$\pm$0.036\\
& &  &  & 9 & 1.870$\pm$0.032\\
& &  &  & 11 & 1.704$\pm$0.033\\
& &  &  &  14& 1.592$\pm$0.055\\
& &  &  &  17& 1.185$\pm$0.023\\
& &  &  &  20& 1.053$\pm$0.027\\
& &  &  &  23& 0.980$\pm$0.020\\
\hline 
2022 Jan 24$^c$ &VLA & VLITE & 1199 & 0.34 & <5.2\\
 & & 21B-360&& 1.12 & 2.285$\pm$0.301\\
& &  &  & 1.37 & 4.466$\pm$0.178\\
& &  &  & 1.62 & 5.132$\pm$0.121\\
& &  &  & 1.88 & 5.239$\pm$0.107\\
& &  &  & 2.5 & 5.343$\pm$0.090\\
& &  &  & 3.5 & 4.649$\pm$0.042\\
& &  &  & 5 &4.120$\pm$0.028\\
& &  &  &  7&3.449$\pm$0.041\\
& &  &  & 9 & 3.056$\pm$0.054\\
& &  &  & 11 & 2.755$\pm$0.057\\
& &  &  &  14& 2.556$\pm$0.041\\
& &  &  &  17& 2.279$\pm$0.048\\
& &  &  &  20& 1.974$\pm$0.042\\
& &  &  &  23& 1.726$\pm$0.023\\
\hline
2022 Mar 17$^c$ &VLA & VLITE & 1251 & 0.34 & <4.0\\
 & & 22A-458&& 1.12 & 4.360$\pm$0.155\\
& &  &  & 1.37 & 5.017$\pm$0.115\\
& &  &  & 1.62 & 5.805$\pm$0.099\\
& &  &  & 1.88 & 6.219$\pm$0.082\\
& &  &  & 2.5 & 6.356$\pm$0.095\\
& &  &  & 3.5 & 6.137$\pm$0.118\\
& &  &  & 5 &5.390$\pm$0.130\\
& &  &  &  7&4.553$\pm$0.140\\
& &  &  & 9 & 4.249$\pm$0.124\\
& &  &  & 11 & 3.313$\pm$0.076\\
& &  &  &  14& 2.450$\pm$0.036\\
& &  &  &  17& 2.735$\pm$0.089\\
& &  &  &  20& 2.478$\pm$0.053\\
& &  &  &  23& 2.166$\pm$0.070\\
\hline
2022 July 11$^d$ &VLA & 22A-458&& 1.12 & 6.927$\pm$0.261\\
 & & && 1.37 & 9.081$\pm$0.141\\
& &  &  & 1.62 & 9.986$\pm$0.146\\
& &  &  & 1.88 & 11.137$\pm$0.144\\
& &  &  & 2.5 & 9.500$\pm$0.771\\
& &  &  & 3.5 & 9.036$\pm$0.862\\
& &  &  & 5 &7.707$\pm$0.200\\
& &  &  &  7&4.830$\pm$0.642\\
\hline
2022 July 12&ALMA & 2021.1.01210.T & 1367 &  97.5& 2.172$\pm$0.061 \\
& & &  &  240& 1.11$\pm$0.04\\
\hline
2022 Sept 11 &MeerKAT & DDT-20220414-YC-01 & 1422 & 0.82 & 5.040$\pm$0.036\\
& & &  &  1.3& 8.697$\pm$0.048\\
\hline
2022 Sept 3 &ATCA & C3472 & 1421 & 2.1 & 10.0$\pm$0.1 \\
& & &  &  5.5& 10.4$\pm$0.2\\
& & &  &  9.0& 8.6$\pm$0.1\\
& & &  &  17& 5.3$\pm$0.2\\
& & &  &  19& 5.0$\pm$0.3\\
\hline
2022 Sept 24 &ALMA & 2021.1.01210.T & 1421 &  97.5& 3.035$\pm$0.33\\
\hline
2022 Dec 19 &VLA & 22B-205& 1528& 1.25 & 8.180$\pm$0.199\\
& &  &  & 1.75 & 12.024$\pm$0.087\\
& &  &  & 2.5 & 13.747$\pm$0.092\\
& &  &  & 3.5 & 13.918$\pm$0.073\\
& &  &  & 5 &13.483$\pm$0.093\\
& &  &  &  7&12.547$\pm$0.124\\
& &  &  & 9 & 11.585$\pm$0.104\\
& &  &  & 11 & 10.578$\pm$0.134\\
& &  &  &  14& 9.953$\pm$0.200\\
& &  &  &  17& 9.153$\pm$0.349\\
& &  &  &  20& 8.892$\pm$0.062\\
& &  &  &  23& 8.462$\pm$0.080\\
\hline
2023 Jan 4 &MeerKAT & SCI-20220822-YC-01 & 1543 & 0.82 & 6.6885$\pm$0.037\\
\hline
2023 Jul 24 &VLA & 23A-241& 1745& 1.25 & 14.307$\pm$0.142\\
& &  &  & 1.75 & 21.094$\pm$0.225\\
& &  &  & 2.5 & 22.484$\pm$0.605\\
& &  &  & 3.5 & 19.259$\pm$0.900\\
& &  &  & 5 &18.873$\pm$0.433\\
& &  &  &  7&14.802$\pm$0.785\\
& &  &  & 9 & 13.416$\pm$0.363\\
& &  &  & 11 & 10.283$\pm$0.396\\
\hline
2023 Sept 22 &VLA & 23A-241& 1804& 1.25 & 13.740$\pm$0.520\\
& &  &  & 1.75 & 21.409$\pm$0.417\\
& &  &  & 2.5 & 28.429$\pm$0.469\\
& &  &  & 3.5 & 26.823$\pm$0.095\\
& &  &  & 6 &22.951$\pm$0.460\\
& &  &  & 9 & 19.506$\pm$0.366\\
& &  &  & 11 & 17.918$\pm$0.413\\
\hline
2023 Dec 8 &VLA & 23B-056& 1881& 1.25 & 16.271$\pm$0.248\\
& &  &  & 1.75 & 21.972$\pm$0.249\\
& &  &  & 2.5 & 26.535$\pm$0.101\\
& &  &  & 3.5 & 26.823$\pm$0.068\\
& &  &  & 5 &26.141$\pm$0.041\\
& &  &  &  7&24.425$\pm$0.071\\
& &  &  & 9 & 22.601$\pm$0.191\\
& &  &  & 11 & 20.879$\pm$0.374\\
& &  &  &  15.5& 17.555$\pm$0.249\\
& &  &  &  17& 16.961$\pm$0.168\\
& &  &  &  20& 16.787$\pm$0.184\\
& &  &  &  23& 16.407$\pm$0.218\\
\hline
2024 Feb 16 &MeerKAT & SCI-20230907-YC-01 & 1951 & 0.82 & 10.949$\pm$0.041\\
\hline
2024 Mar 10 &VLA & SC240292& 1974& 1.25 & 16.714$\pm$0.854\\
& &  &  & 1.75 & 25.717$\pm$0.150\\
& &  &  & 2.5 & 28.468$\pm$0.181\\
& &  &  & 3.5 & 30.190$\pm$0.075\\
& &  &  & 5 &29.604$\pm$0.119\\
& &  &  &  7&27.874$\pm$0.178\\
& &  &  & 9 & 25.810$\pm$0.180\\
& &  &  & 11 & 24.154$\pm$0.209\\
& &  &  &  14& 22.809$\pm$0.191\\
& &  &  &  17& 21.417$\pm$0.248\\
& &  &  &  20& 20.226$\pm$0.115\\
& &  &  &  23& 19.324$\pm$0.120\\
\hline
2024 June 7 &VLA & 24A-353& 2063& 1.25 & 22.177$\pm$0.174\\
& &  &  & 1.75 & 29.723$\pm$0.162\\
& &  &  & 2.5 & 37.286$\pm$0.266\\
& &  &  & 3.5 & 37.104$\pm$0.228\\
& &  &  & 5 &32.388$\pm$0.334\\
& &  &  &  7&28.884$\pm$0.490\\
& &  &  & 9 & 27.736$\pm$0.291\\
& &  &  & 11 & 25.523$\pm$0.465\\
& &  &  &  14& 24.589$\pm$0.351\\
& &  &  &  17& 22.984$\pm$0.238\\
& &  &  &  20& 20.905$\pm$0.309\\
& &  &  &  23& 17.644$\pm$0.548\\
\hline
2024 Sept 12 &VLA & 24A-353& 2160& 1.25 & 22.190$\pm$0.053\\
& &  &  & 1.75 & 29.094$\pm$0.330\\
& &  &  & 2.5 & 33.810$\pm$0.257\\
& &  &  & 3.5 & 36.357$\pm$0.240\\
& &  &  & 5 &33.993$\pm$0.474\\
& &  &  &  7&32.031$\pm$0.566\\
& &  &  & 9 & 29.742$\pm$0.345\\
& &  &  & 11 & 26.878$\pm$0.366\\
& &  &  &  14& 24.377$\pm$0.348\\
& &  &  &  17& 20.218$\pm$0.332\\
& &  &  &  20& 21.693$\pm$1.014\\
& &  &  & 23& 18.173$\pm$1.072\\
\hline
2024 Sept 27 &MeerKAT & SCI-20230907-YC-01 & 2175 & 0.82 & 12.776$\pm$0.053\\
\hline
2024 Oct 20 &SMA & POETS & 2198 & 225.5 & 6.32$\pm$0.31 \\
\enddata
\tablecomments{$^a$ These values are measured relative to the time of optical discovery, 2018 Oct 14.\\ $^b$ Limits are $3\sigma$.\\ $^c$ VLA observations with updated flux measurements from \citet{p1} due to pipeline error; see Section \ref{sec:obs-radio}.  We note explicitly that we do not report observations from \citet{p1} that were unaffected by this error, and those are reported in \citet{p1}. \\$^d$ Observation taken in API conditions adverse to higher frequency observations.  We exclude higher bands affected.}
\end{deluxetable*}

\bibliography{sample631}{}
\bibliographystyle{aasjournal}

\end{document}